\documentclass[useAMS,usenatbib]{mn2e}

\usepackage{graphicx}
\usepackage{amssymb}
\title[GRB081028]{GRB\,081028 and its late-time afterglow re-brightening}
\author[Margutti et al.]{R. Margutti$^{1,2}$\thanks{E-mail:
raffaella.margutti@brera.inaf.it (RM)}, F. Genet$^{3}$, J. Granot$^{3}$, R. Barniol Duran$^{4}$, 
C. Guidorzi$^{5,2}$, \and G. Chincarini$^{1,2}$, J. Mao$^{2,13}$, P. Schady$^{6}$, T. Sakamoto$^{7}$, A. A. Miller$^{8}$, G. 
Olofsson$^{9}$, \and J.S. Bloom$^{8}$, P.A. Evans$^{10}$, J.P.U. Fynbo$^{11}$, D. Malesani$^{11}$, A. Moretti$^{2}$,
F. Pasotti$^{2}$, \and D. Starr$^{8}$, D.N. Burrows$^{12}$, S.D. Barthelmy$^{7}$, P.W.A. Roming$^{12}$, N. Gehrels$^{7}$ \\
$^{1}$Universit\`a degli studi Milano Bicocca, P.za della Scienza 3, Milano 20126, Italy\\
$^{2}$INAF Osservatorio Astronomico di Brera, via Bianchi 46, Merate 23807, Italy \\
$^{3}$Centre for Astrophysics Research, University of Hertfordshire, UK\\
$^{4}$Department of Physics, University of Texas at Austin, Austin, TX 78712, USA\\
$^{5}$Dipartimento di Fisica, Universit\`a di Ferrara, via Saragat 1, 44100 Ferrara, Italy\\
$^{6}$The UCL Mullard Space Science Laboratory, Holmbury St Mary, Dorking, Surrey RH5 6NT\\
$^{9}$Stockholm Observatory, Stockholm University, Astronomy Department AlbaNova Research Center, 106 91 Stockholm, Sweden\\
$^{11}$Dark Cosmology Centre, Niels Bohr Institute, University of C\o penhagen, Juliane MariesVej 30, 2100 Copenhagen O, Denmark\\
$^{8}$Department of Astronomy, University of California, Berkeley, CA 94720-3411, USA\\
$^{10}$X-ray and Observational Astronomy Group, Department of Physics and Astronomy, University of Leicester, LE1 7RH, UK\\
$^{12}$Department of Astronomy and Astrophysics, Pennsylvania State University, 525 Davey Lab, University Park, PA 16802\\
$^{7}$NASA Goddard Space Flight Center, Greenbelt, MD 20771\\ 
$^{13}$Yunnan  Observatory,  Chinese Academy of Sciences, P.O. Box 110, Kunming, Yunnan  Province, China\\}
\begin{document}

\date{Accepted 200? Month Day  Received 2009  Month Day; in original form 2009 March 19}

\pagerange{\pageref{firstpage}--\pageref{lastpage}} \pubyear{2009}

\maketitle

\label{firstpage}

\begin{abstract}
\emph{Swift} captured for the first time a smoothly rising X-ray re-brightening
of clear non-flaring origin after the steep decay in a long gamma-ray burst (GRB): 
GRB\,081028. A rising phase is likely present in all GRBs but is usually hidden by
the prompt tail emission and constitutes the first manifestation of what is later 
to give rise to the shallow decay phase. Contemporaneous optical observations reveal 
a rapid evolution of the injection frequency of a fast cooling synchrotron spectrum 
through the optical band, which disfavours 
the afterglow onset (start of the forward shock emission along our line of sight when 
the outflow is decelerated) as the origin of the observed re-brightening.
We investigate alternative scenarios and find that the observations are 
consistent with the predictions for a narrow jet viewed off-axis.
The high on-axis energy 
budget implied by this interpretation suggests different physical origins of the prompt
and (late) afterglow emission.
Strong spectral softening takes place from the prompt to the steep 
decay phase: we track the evolution of the spectral peak energy from
the $\gamma$-rays to the X-rays and highlight the problems of 
 the high latitude and adiabatic cooling interpretations.
Notably, a softening of both the high and low spectral slopes with time is also observed. 
We discuss the low on-axis radiative efficiency of GRB\,081028 comparing its properties 
against a sample of \emph{Swift} long GRBs with secure $E_{\rm{\gamma,iso}}$ measurements.
\end{abstract}


\section{Introduction}
Gamma ray bursts (GRBs) are transient events able to outshine 
the $\gamma$-ray sky for a few seconds to a few minutes. The discovery of their
optical \citep{Paradijs97} and X-ray \citep{Costa97} long-lasting 
counterparts represented a breakthrough for GRB science. Unfortunately, due
to technological limitations, the X-ray observations were able to track the afterglow 
evolution starting hours after the trigger: only 
after the launch of the \emph{Swift} satellite in 2004 \citep{Gehrels04} 
was this gap between 
the end of the prompt emission and several hours after the onset 
of the explosion filled with X-ray observations. A canonical
picture was then established (see e.g., \citealt{Nousek06}), with 
four different stages describing the overall structure of the
X-ray afterglows: an initial steep decay, a shallow-decay 
phase, a normal decay and a jet-like decay stage. Erratic flares
are found to be superimposed mainly to the first and second stage of emission.
An interesting possibility is that the four light-curve phases instead belong
to only two different components of emission (see e.g., \citealt{Willingale07}):
the first, connected to the activity of the central engine giving rise
to the prompt emission, comprises the flares (\citealt{Chincarini07} and
references therein) and the steep-decay phase;
the second is instead related to the interaction of the outflow with the 
external medium and manifests itself in the X-ray regime through the 
shallow, normal and jet-like decay. Observations able
to further characterise the two components are therefore of particular 
interest.

The smooth connection of the X-ray steep decay light-curve phase with the 
prompt $\gamma$-ray emission strongly suggests a common physical origin
(\citealt{Tagliaferri05}; \citealt{Obrien06}): the high latitude 
emission (HLE) model (\citealt{Fenimore96}; \citealt{Kumar00}) predicts
that steep decay photons originate from the delay in the arrival time
of prompt emission photons due to the longer path length from larger angles
relative to our line of sight, giving rise to the
$\alpha=\beta+2$ relation (where $\alpha$ is the light-curve
decay index and $\beta$ is the spectral energy index). No spectral evolution
is expected in the simplest formulation of the HLE effect in the
case of a simple power-law prompt spectrum. Observations
say the opposite: significant variations of the photon index have been
found in the majority of GRBs during the steep decay phase 
(see e.g., \citealt{Zhang07}); more than this, the absorbed simple 
power-law (SPL) has proved to be a poor description of the spectral energy 
distribution of the steep decay phase for GRBs with the best statistics\footnote
{The limited $0.3 -10 \,\rm{keV}$ spectral 
coverage of the Swift X-Ray Telescope, XRT \citep{Burrows05}, and the degeneracy 
between  the variables of the spectral fit can in principle lead to the
identification of an SPL behaviour in intrinsically non-SPL spectra with
poor statistics.}.
A careful analysis of these events has shown their spectra to be best 
fit by an evolving Band function \citep{Band93}, establishing the link 
between steep decay and prompt emission photons also from the spectral 
point of view (see e.g., GRB\,060614, \citealt{Mangano07}; GRB\,070616, 
\citealt{Starling08}): caused by the shift of the 
Band spectrum, a temporal steep decay phase and a spectral softening 
appear simultaneously (see e.g. \citealt{Zhang09}, \citealt{Qin09}).
In particular, the peak energy of the $\nu F_{\nu}$ spectrum is found to 
evolve to lower values, from the $\gamma$-ray to the soft X-ray 
energy range. Both the low (as observed for GRB\,070616) and the high-energy 
portion of the spectrum are likely to soften with time, but 
no observation is reported to confirm the high energy index behaviour 
during the prompt and steep decay phase.
The observed spectral evolution with time is an invaluable footprint
of the physical mechanisms at work: observations able
to constrain the behaviour of the spectral parameters with time are 
therefore of primary importance.

By contrast, no spectral evolution 
is observed in the X-ray during the
shallow decay phase (see e.g. \citealt{Liang07}) experienced by most GRBs between
$\sim 10^2$ s and $10^3 -10^ 4$ s. An unexpected discovery of the 
\emph{Swift} mission, the shallow decay is the first light-curve 
phase linked to the second emission component. A variety of 
theoretical explanations have been put forward. The proposed 
models include: energy injection (\citealt{Panaitescu06}; \citealt{Rees98}; 
\citealt{Granot06}; \citealt{Zhang06}); 
reverse shock (see e.g., \citealt{Genet07}); 
time dependent micro-physical parameters (see e.g. \citealt{Granot06b};
\citealt{Ioka06});
off-axis emission \citep{Eichler06}; dust scattering
\citep{Shao07}. 
The predictions of all these models can only be compared 
to observations tracking the flat and decay phase of the
second emission component, since its  rise is usually missed
in the X-ray regime, being hidden by the tail of the prompt emission.

GRB\,081028 is the first and unique event for which \emph{Swift}
was able to capture the rise of the second emission component\footnote
{There are a handful of long GRBs detected by \emph{Swift}
with a possible X-ray rise of non-flaring origin. Among them:
GRB\,070328, \cite{Markwardt07}; GRB\,080229A, \cite{Cannizzo08};
GRB\,080307, \cite{Page09}
(see \citealt{Page09} and references therein). 
However, in none of these cases has an X-ray 
steep decay been observed.
A smooth rise in the X-rays has been observed in the short GRB\,050724.}:
the time properties of its rising phase can be constrained for the 
first time while contemporaneous optical observations 
allow us to track the evolution of a break energy of the spectrum
through the optical band. GRB\,081028 is also one of the lucky cases
showing a spectrally evolving prompt emission where the evolution
of the spectral parameters can be studied from $\gamma$-rays to
X-rays, from the trigger time to $\sim 1000$ s. A hard to soft spectral 
evolution is clearly taking place beginning with the prompt emission 
and extending to the steep decay phase, as already found for other 
\emph{Swift} GRBs (GRB\,060614, \citealt{Mangano07}; GRB\,070616, 
\citealt{Starling08}, are showcases
in this respect). Notably, for GRB\,081028 a softening  of the slope of a Band 
function  \citep{Band93} above $E_{\rm p}$ is also observed.

The paper is organised as follows: \emph{Swift} and ground-based 
observations are described in Sect. \ref{sec:observations}; data 
reduction and preliminary analysis are reported in Sect. 
\ref{Sec:datared}, while in Sect. \ref{sec:analysis} the results
of a detailed spectral and temporal multi-wavelength analysis 
are outlined and discussed in Sect. \ref{sec:discussion}.
Conclusions are drawn in Sect. \ref{sec:conclusion}.

The phenomenology of the burst is presented in the observer frame unless 
otherwise stated. The convention $F_{\nu}(\nu,t)\propto \nu^{-\beta}t^{-\alpha}$
is followed, where $\beta$ is the spectral energy index, related to the 
spectral photon
index $\Gamma$ by $\Gamma=\beta+1$.
All the quoted uncertainties are given at 68\% confidence
level (c.l.): a warning is added if it is not the case.
The convention 
$Q_{x}=Q/10^{x}$ has been adopted in cgs units unless otherwise stated.
Standard cosmological quantities have been adopted: 
$H_{0}=70\,\rm{km\,s^{-1}\,Mpc^{-1}}$, $\Omega_{\Lambda }=0.7$, 
$\Omega_{\rm{M}}=0.3$.\\

\section{Observations}
\label{sec:observations}
GRB\,081028 triggered the \emph{Swift} Burst Alert Telescope (BAT; \citealt{Barthelmy05}) 
on 2008-10-28 at 00:25:00 UT \citep{Guidorzi08}. The spacecraft immediately slewed to the 
burst allowing the X-ray Telescope (XRT; \citealt{Burrows05}) to collect 
photons starting at $T+191\,\rm{s}$ after the trigger: a bright and fading X-ray 
afterglow was discovered. The UV/Optical Telescope (UVOT, \citealt{Roming05})
began observing at $T+210\,\rm{s}$. In the first orbit of observations, no 
afterglow candidate was detected in any of the UVOT filters in either the individual 
or co-added exposures. A careful re-analysis of the acquired data
revealed the presence of a source with a White band magnitude of $20.9$ at
$\sim T+270\,\rm{s}$ (this paper). A refined position was quickly available 
thanks to the XRT-UVOT alignment procedure and the match of UVOT field 
sources to the USNO-B1 catalogue (see \citealt{Goad07}  for details): 
R.A.(J2000)=$08^{\rm{h}}07^{\rm{m}}34.76^{\rm{s}}$, 
Dec.(J2000)=$+02^{\circ}18'29.8\arcsec$ with a 90\% error radius of 1.5 arcsec \citep{Evans08}.  
Starting at $\sim T+9\,\rm{ks}$ the X-ray light-curve shows a remarkable re-brightening
\citep{Guidorzi08b}, see Fig. \ref{Fig:plottot_lc}: this was later detected in 
ground-based near-infrared (NIR) and optical observations. Preliminary analysis results for this burst 
were reported in \cite{Guidorzi08c}.

The Telescope a Action Rapide pour les Objets Transitoires (TAROT; 
\citealt{Klotz08}) began observing $566.4\,\rm{s}$ after the trigger
under poor weather conditions: no variable source was detected down to
$R\sim17.4$.

The optical afterglow was discovered by the Gamma-Ray Burst Optical and Near-Infrared
Detector (GROND; \citealt{Greiner08}). The observations started 20.9 ks after the 
trigger: the afterglow was simultaneously detected in the $g'r'i'z'JHK$ bands 
\citep{Clemens08} with the following preliminary magnitudes: 
$g'= 19.9\pm{0.1}$; $r'=19.3\pm{0.1}$; $i'=19.2\pm{0.1}$; $z'=19.1\pm{0.1}$;
$J=19.0\pm{0.15}$; $H=18.7\pm{0.15}$; $K=19.0\pm{0.15}$, with a net exposure
of $264$ and $240\,\rm{s}$ for the $g'r'i'z'$ and the $JHK$ bands 
respectively. Further GROND observations were reported by \cite{Clemens08b}
$113\,\rm{ks}$ after the trigger with $460\,\rm{s}$ of total exposures in 
$g'r'i'z'$ and $480\,\rm{s}$ in $JHK$. Preliminary magnitudes are reported below:
$g'= 21.26\pm{0.05}$; $r'=20.49\pm{0.05}$; $i'=20.24\pm{0.05}$; $z'=19.99\pm{0.05}$;
$J=19.6\pm{0.1}$. The source showed a clear 
fading with respect to the first epoch, confirming its nature as a GRB afterglow.

The Nordic Optical Telescope (NOT) imaged the field of GRB\,081028 $\sim6\,\rm{hr}$
after the trigger and independently confirmed the optical afterglow with a
magnitude $R\sim19.2$ \citep{Olofsson08}. Because of the very poor 
sky conditions only 519 frames out of 9000 could be used, with a total exposure 
of 51.9 s. The average time for the observations is estimated to be 05:53:00 UT.
Image reduction was carried out by following standard procedures.

An UV/optical re-brightening was discovered by the UVOT starting 
$T+10\,\rm{ks}$, simultaneous to the X-ray re-brightening. The afterglow 
was detected in the $v$, $b$ and $u$-band filters \citep{Shady08}. 
The UVOT photometric data-set of GRB\,081028 is reported in Tab.
\ref{Tab:UVOTdata}. We refer to \cite{Poole08} for a detailed description
of the UVOT photometric system.

The rising optical afterglow was independently confirmed by the Crimean telescope
for Asteroid Observations (CrAO) and by the Peters Automated Infrared 
Imaging Telescope (PAIRITEL; \citealt{Bloom06}). CrAO observations were carried out starting at
$\sim T+1\,\rm{ks}$ and revealed a sharp rising optical afterglow peaking after
$T+9.4\,\rm{ks}$: $R=21.62\pm0.07$ at $t=T+1.8\,\rm{ks}$; $I=21.32\pm0.09$ at 
$t=T+3.6\,\rm{ks}$; $I=21.43\pm0.09$ at $t=T+5.5\,\rm{ks}$; $I=21.20\pm0.08$ at 
$t=T+7.5\,\rm{ks}$; $I=20.66\pm0.05$ at $t=T+9.4\,\rm{ks}$ \citep{Rumyantsev08}.

PAIRITEL observations were carried out $40\,\rm{ks}$ after the trigger: the 
afterglow was simultaneously detected in the $J$, $H$, and $K_{\rm{s}}$ filters with 
a preliminary photometry $J=17.7\pm0.1$, $H =17.0\pm0.1$ and  $K_{\rm{s}} = 16.1\pm 0.1$
\citep{Miller08}. A total of 472 individual 7.8 s exposures were obtained under bad conditions 
(seeing $\ga$ 3$''$) for a total exposure time of $\sim$3682 s. 
The data were reduced and analysed using the standard PAIRITEL pipeline
\citep{Bloom06}. Photometry calibration was done against the 2MASS system.
The resulting fluxes and magnitudes are consistent with the values 
reported by \cite{Miller08}: however, this work should be
considered to supersede the previous findings.
The ground-based photometric data-set of GRB\,081028 is reported in Tab.
\ref{Tab:OpticalData}  while the photometric optical observations of GRB\,081028 are 
portrayed in Fig. \ref{Fig:plottot_lc}.

A spectrum of the GRB\,081028 afterglow was taken with the Magellan Echellette Spectrograph (MagE) 
on the Magellan/Clay 6.5-m telescope at $\sim T+27\,\rm{ks}$ for a total
integration time of $1.8\,\rm{ks}$. The identification of absorption features
including SII, NV, SiIV, CIV and FeII  allowed the measurement of the redshift $z=3.038$
together with the discovery of several intervening absorbers \citep{Berger08}.

According to \cite{Schlegel98} the Galactic reddening along the line of sight
of GRB\,081028 is $E(B-V)=0.03$. 
\section{Swift Data Reduction and preliminary analysis}
\label{Sec:datared}
\label{SubSec:SwiftXRTdata}

\begin{figure}
\vskip -0.0 true cm
\centering
    \includegraphics[scale=0.65]{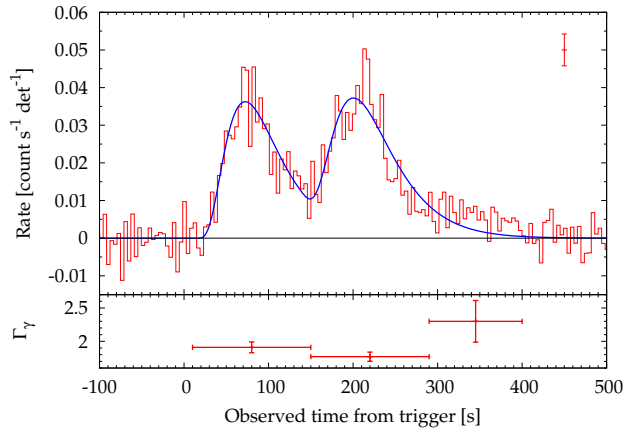}
      \caption{Top panel: BAT 15-150 keV mask weighted light-curve (binning 
        time of 4.096 s). Solid blue line: $15-150\,\rm{keV}$ light-curve 
        best fit using Norris et al. (2005) profiles. The typical $1~\sigma$ error size 
        is also shown. Bottom panel: best fit photon index $\Gamma_{\gamma}$
        as a function of time (errors are provided at the 90\% c.l.).}
\label{Fig:BATplindex}
\end{figure}

The BAT data have been processed using standard Swift-BAT analysis tools  
within \textsc{heasoft} (v.6.6.1).  The ground-refined coordinates 
provided by \cite{Barthelmy08} have been adopted in the following analysis.
Standard filtering and screening criteria have been applied.
The mask-weighted background subtracted 
$15-150\,\rm{keV}$ is shown in Fig. \ref{Fig:BATplindex}, top panel. 
The mask-weighting procedure is also applied to produce weighted, 
background subtracted counts spectra.
Before fitting the spectra, we group the 
energy channels requiring a 3-$\sigma$ threshold on each group;
the threshold has been lowered to 2-$\sigma$ for spectra with 
poor statistics. The spectra are fit within \textsc{Xspec} v.12.5
with a simple power-law with pegged normalisation (\textsc{pegpwrlw}).
The best fit photon indices resulting from this procedure are shown
in Fig. \ref{Fig:BATplindex}, bottom panel.


\begin{figure*}
\centering
\includegraphics[scale=0.7]{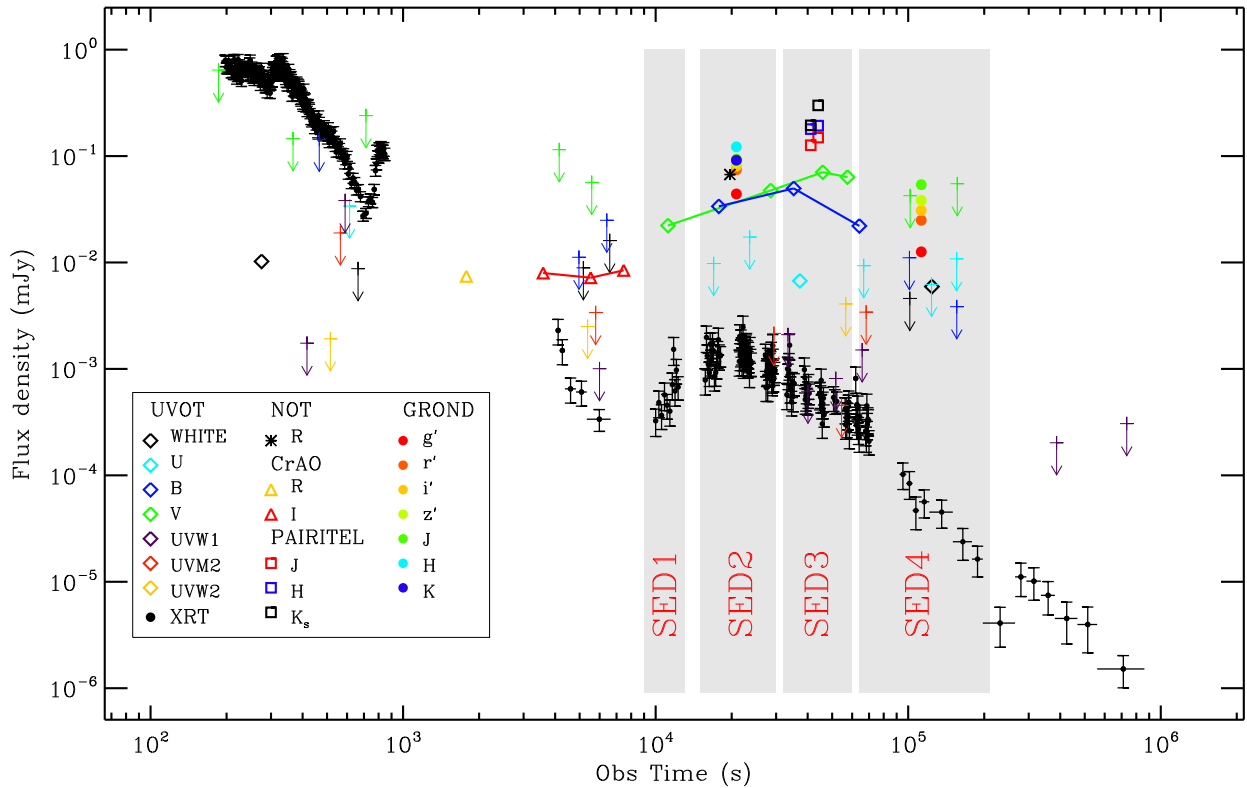}
\caption{Complete data set for GRB\,081028 
starting 200 s after the trigger including X-ray (XRT, flux density estimated at 1 keV), 
UV/visible/NIR (UVOT, GROND, PAIRITEL, CrAO, NOT) observations.
The arrows indicate 3-$\sigma$ upper limits of UVOT observations.
The shaded regions indicate the time intervals of 
extraction of the SEDs.}
\label{Fig:plottot_lc}
\end{figure*}

XRT data have also been processed with  \textsc{heasoft}
(v. 6.6.1) and corresponding
calibration files: standard filtering and screening criteria have been applied.
The first orbit data were acquired entirely in WT mode reaching a maximum count rate $\sim140
\,\rm{counts\,s^{-1}}$.  We apply standard pile-up corrections following
the prescriptions of  \cite{Romano06} when necessary. Starting from $\sim
10\,\rm{ks}$ \emph{Swift}-XRT switched to PC mode to follow the fading
of the source:  events are then extracted using different region shapes
and sizes in order to maximize the signal-to-noise (SN) ratio. The background
is estimated from a source free portion of the sky. The resulting X-ray light-curve
is shown in Fig. \ref{Fig:plottot_lc}: the displayed data binning assures 
a minimum SN equals to 4 (10) for PC (WT) data. In this way the strong
variability of WT data can be fully appreciated without losing 
information on the late time behaviour. We perform automatic time resolved spectral 
analysis, accumulating signal
over time intervals defined to contain a minimum of $\sim2000$ photons each.
The spectral channels have been grouped to provide a minimum of 20 counts per bin. 
The Galactic column density in the direction of the burst is estimated to be 
$3.96\times10^{20}\,\rm{cm^{-2}}$ (weighted average value from
the \citealt{Kalberla05} map). Spectral fitting is done within \textsc{Xspec} 
(v.12.5) using a photo-electrically absorbed simple power law (SPL) model.
The Galactic 
absorption component is frozen at the Galactic value together with the redshift,
while we leave the intrinsic column density free to vary during the first run of 
the program. A count-to-flux conversion factor is worked out from the best
fit model for each time interval for which we are able to extract a spectrum.
This value is considered reliable if the respective $\chi^2/\rm{dof}$ (chi-square over
degrees of freedom) implies a P-value (probability of obtaining a result at least as 
extreme as the one that is actually observed) higher than 5\%.
The discrete set of reliable count-to-flux conversion factors is then used to 
produce a continuous count-to-flux conversion factor through interpolation. 
This procedure produces flux and luminosity light-curves where the possible 
spectral evolution of the source is properly taken into account 
(Fig. \ref{Fig:plottot_lc}). 
In the case of GRB\,081028 this is particularly important: 
the simple power law photon index evolves from $\Gamma\sim1.2$ to $\Gamma\sim3$ 
during the steep decay phase (Fig. \ref{Fig:PhotonIndex_nhtot}), inducing a variation of 
a multiplicative factor $\sim1.7$ in the count-to-flux conversion factor.
As a second run, we remove one degree of freedom from the spectral fitting
procedure, noting the absence of spectral evolution during the X-ray re-brightening 
in the X-ray regime
(see Sect. \ref{SubSec:specXRT}). This gives the possibility to obtain a reliable
estimate of the intrinsic neutral Hydrogen column density $N_{\rm{H,z}}$ of 
GRB\,081028: the PC spectrum accumulated over the time interval 
$10-652\,\rm{ks}$ can be adequately fit by an absorbed SPL model with best fit
photon index $\Gamma=2.09\pm0.07$ and 
$N_{\rm{H,z}}=(0.52\pm0.25)\times10^{22}\,\rm{cm^{-2}}$ ($90\%$ c.l. uncertainties are provided). 
The flux-luminosity calibration procedure is then re-run freezing the intrinsic absorption
component to this value.
 
The UVOT photometry was performed using standard tools
\citep{Poole08} and is detailed in Tab. \ref{Tab:UVOTdata}.

\section{Analysis and results}
\label{sec:analysis}
\subsection{Temporal analysis of BAT (15-150 keV) data}
\label{SubSec:taBAT}

\begin{table}\footnotesize
\begin{center}
\begin{tabular}{l|cc}
\hline
& Pulse 1&  Pulse 2\\
\hline
 $t_{\rm{peak}}$ (s)& $72.3\pm3.5$ & $202.7\pm3.3$\\
 $t_{\rm{s}}$ (s)&$5.4\pm17.5$ &$125.6\pm18.1$\\
 $t_{\rm{rise}}$ (s)& $32.6\pm3.7$&$36.4\pm4.1$\\
 $t_{\rm{decay}}$ (s)& $63.4\pm8.1$&$70.0\pm5.2$\\
 $w$ (s)& $96.0\pm7.9$&$105.4\pm6.2$\\
 $k$& $0.32\pm0.09$&$0.31\pm0.07$\\
 $A$ ($\rm{count\,\,\rm{s^{-1}}}\,\rm{det^{-1}}$)& $(3.6 \pm0.2)10^{-2}$ &$(3.5 \pm0.2)10^{-2}$\\
 Fluence ($\rm{erg\,\,cm^{-2}}$) & $(1.81\pm0.14)10^{-6}$ & $(1.83\pm0.11)10^{-6}$\\
 \hline
 $\chi^{2}/\rm{dof} $& \multicolumn{2}{c}{171/114}\\
\hline

\end{tabular}
\caption{Best fit parameters and related quantities resulting
from the modelling of the prompt 15-150 keV emission with two
Norris et al. (2005) profiles. From top to bottom: peak time,
start time, $1/e$ rise time, $1/e$ decay time, $1/e$ pulse 
width, pulse asymmetry, peak count-rate and statistical information.
The $\chi^2$ value mainly reflects the partial failure of the
fitting function to adequately model the peaks of the pulses 
(see Norris et al. 2005 for details).}
\label{Tab:BATta}
\end{center}
\end{table}

The mask-weighted light-curve consists of two main pulses peaking at 
$T+70\,\rm{s}$ and $\sim T+200\,\rm{s}$ followed by a long lasting 
tail out to $\sim T+400\,\rm{s}$. 
In the time interval $T-100\,\rm{s}$ $T+400\,\rm{s}$, the light-curve
can be fit by a combination of two \cite{Norris05} profiles 
(Fig. \ref{Fig:BATplindex}, top panel), each profile consisting 
of the inverse of the product of two exponentials, one increasing
and one decreasing with time. The best fit parameters and
related quantities are reported in Table \ref{Tab:BATta}: the parameters
are defined following \cite{Norris05}; we account for the entire
covariance matrix during the error propagation procedure. The GRB
prompt signal has a $T_{90}$ duration of $261.0\pm28.7$ s and a $T_{50}= 128.2 \pm 7.7 $ s.

The temporal variability of this burst has been characterised in two 
different ways. First, following \cite{Rizzuto07} we compute
a variability measure 
$\rm{Var}(15-150\,\rm{keV})=(5.0\pm0.14)\times 10^{-2}$. Second, 
we adopt the power spectrum analysis in the time domain
(\citealt{Li01}; \citealt{Li02}): unlike the 
Fourier spectrum, this is suitable to study the rms variations at 
different time-scales. See \cite{Margutti08} and Margutti et al. in 
prep. for details about the application of this technique to the GRB 
prompt emission. In particular, we define the fractional power density
(fpd) as the ratio between the temporal power of the source signal
and the mean count rate 
squared. This quantity is demonstrated to show a peak at the 
characteristic time scales of variability of the signal. We assess the 
significance of each fpd peak via Montecarlo simulations.
The fpd of GRB\,081028 shows a clear peak around 70 s (time scale related
to the width of the two \citealt{Norris05} profiles). Below 70 s the 
fpd shows a first peak at $\Delta t\sim2\,\rm{s}$ and then a second peak at  
$\Delta t\sim6\,\rm{s}$, both at 1-$\sigma$ c.l. The signal shows power 
in excess of the noise at 2-$\sigma$ c.l. significance for time scales
$\Delta t\geq 32\,\rm{s}$. 

\subsection{Spectral analysis of BAT (15-150 keV) data}
\label{SubSec:specBAT}
\begin{table*}\footnotesize
\begin{center}
\begin{tabular}{ccccccccc}
\hline
Interval & Model &$t_{\rm{start}}$ & $t_{\rm{stop}}$& $\Gamma,\alpha$& $E_{\rm{p}}$& Fluence& $\chi^{2}/\rm{dof}$ & P-value\\
         &       & (s)             &     (s)        &                &   (keV)     & ($\rm{erg\,cm^{-2}}$) & &\\
\hline
$T_{90}$ & Pl    &52.9             &   317.2        &$1.82\pm0.09$   &   --         &$(3.3\pm0.20)\times 10^{-6}$& $31.8/31$&$43\%$\\
         & Cutpl &52.9             &   317.2        &$1.3\pm0.4$     & $65^{+42}_{-11}$&  $(3.15\pm0.20)\times 10^{-6}$& $25.8/30$& $69\%$\\
\hline
Total    & Pl     & 0.0            &   400.0        & $1.89\pm0.09$& -- & $(3.7\pm0.20)\times 10^{-6}$&$37.4/32$&$23\%$\\
         & Cutpl  & 0.0            &   400.0        & $1.3\pm0.4$&$55^{+20}_{-9}$&$(3.45\pm0.19)\times 10^{-6}$&$30.1/31$&$51\%$\\
\hline
Pulse 1  & Pl     & 10.0           &   150.0        &$1.91\pm0.13$&-- &$(1.60\pm0.12)\times 10^{-6}$&$18.0/24$&$80\%$\\
         & Cutpl  & 10.0           &   150.0        &$1.1\pm0.6$&$49^{+18}_{-9}$&$(1.47\pm0.11)\times 10^{-6}$&$12.0/23$&$97\%$\\
\hline
Pulse 2 & Pl      & 150.0          &   290.0        &$1.77\pm0.11$&-- &$(1.79\pm0.11)\times 10^{-6}$&$33.8/29$&$25\%$\\
        & Cutpl   & 150.0          &   290.0        &$1.22\pm0.45$&$69^{+87}_{-14}$&$(1.47\pm0.11)\times 10^{-6}$&$29.3/28$&$40\%$\\
\hline
 \end{tabular}
\caption{Best fit parameters derived from the spectral modelling of 15-150 keV data
using a power law with pegged normalisation (Pl, \textsc{pegpwrlw} within \textsc{Xspec}) 
and a cut-off power-law model with the peak energy of the $\nu F_{\nu}$ spectrum as free parameter
(Cutpl). From left to right: name of the interval
of the extraction of the spectrum we refer to throughout the paper; spectral model
used; start and stop times of extraction of the spectrum; best fit photon index $\Gamma$
for a Pl model or cutoff power-law index for a Cutpl model; best fit peak energy of 
the $\nu F_{\nu}$ spectrum; fluence; statistical information about the fit. }
\label{Tab:BATspec}
\end{center}
\end{table*}
We extract several spectra in different time intervals and then fit the data using 
different models to better constrain the spectral evolution of GRB\,081028 
in the 15-150 keV energy band.
The first spectrum is extracted during the $T_{90}$ duration of the
burst; a second spectrum is accumulated during the entire duration
of the 15-150 keV emission; finally, the signal between 10 s and 290 s
from trigger has been split into two parts, taking 150 s as dividing time,
to characterise the spectral properties of the two prompt emission pulses.
The resulting spectra are then fit using a simple power-law and a 
cut-off power-law models within \textsc{Xspec}. The results
are reported in Table \ref{Tab:BATspec}. The measured simple power law
photon index around 2 suggests that BAT observed a portion of
an intrinsically Band spectrum \citep{Band93}. Consistent with this
scenario, the cut-off power law model always provides a better fit which 
is able to constrain the peak energy value ($E_{\rm{p}}$, peak of the 
$\nu F_{\nu}$ spectrum) within the BAT energy range. 



The best fit parameters of the cut-off power-law model applied to the
total spectrum of Table \ref{Tab:BATspec} imply 
$E_{\rm{iso,\gamma}}(1-10^4\,\rm{keV})=(1.1\pm0.1)\times 10^{53}\,\rm{erg}$.
The respective rest frame peak energy is 
$E_{\rm{p,i}}=(1+z)E_{\rm{p}}=222^{+81}_{-36}\,\rm{keV}$, placing GRB\,081028
within the 2-$\sigma$ region of the Amati relation \citep{Amati06}.
The burst is characterised by an isotropic $10^2-10^3$ keV (rest frame)
$L_{\rm{iso}}=(2.85\pm0.25)\times10^{51}\,\rm{erg\,s^{-1}}$. This information
together with the variability measure $\rm{Var}(15-150\,\rm{keV})=(5.0\pm0.14)\times 10^{-2}$
makes GRB\,081028 perfectly consistent with the luminosity variability relation
(see \citealt{Reichart01}; \citealt{Guidorzi05}; \citealt{Rizzuto07}).

\subsection{Temporal analysis of XRT (0.3-10 keV) data}
\label{SubSec:taXRT}

\begin{table*}\footnotesize
\begin{center}
\begin{tabular}{rrrrrrrrrrrrr}
\hline
$n_2$& $c$& $n_1$& $a$ & $b$ & $d_1$& $t_{\rm{br}_1}$& $n_3$ & $e$& $d_2$& $t_{\rm{br}_2}$ & $\chi^2/\rm{dof}$\\
     &    &      &     &     &      & (ks)           &       &    &      &   (ks)          &                          \\
\hline
$10^{18.3\pm5.5}$& $-5.2\pm1.5$& $1.2\pm1.1$& $-4.5\pm3.3$& $2.1\pm0.1$& $2.4\pm2.0$& $15.5\pm6.3$& $-$& $-$& $-$& $-$& $164.8/145$\\
$10^{29.9\pm6.5}$& $-7.60\pm1.8$ & $0.31\pm0.02$ & $-1.8\pm0.3$&$1.3\pm0.1$&$0.1$& $19.5\pm0.7$&$0.06$&$2.3\pm0.1$& $0.05$& 62& $147.1/143$\\
\hline
\end{tabular}
\caption{Best-fit parameters of the XRT light-curve modelling starting from 3 ks 
after the trigger. The first
(second) line refers to Eq. \ref{Eq:plBeuermannfree} (Eq. \ref{Eq:rebfin}).}
\label{Tab:XRTrebfit}
\end{center}
\end{table*}

\begin{figure}
\centering
\includegraphics[scale=0.58]{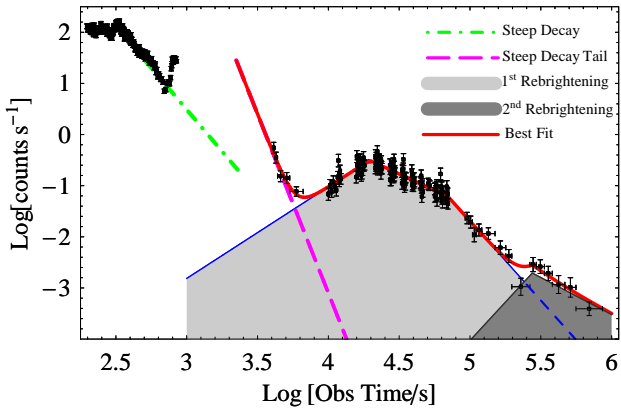}
\caption{0.3-10 keV X-ray afterglow split into different components.
Green dot-dashed line: steep decay; purple long dashed line:
pre-rebrightening component; light grey region: first re-brightening component;
dark grey region: second re-brightening component; red solid line:
best fit model. See Sect. \ref{subsubsec:drop} for details.}
\label{Fig:XRTcomponents}
\end{figure}

\begin{figure}
\centering
\includegraphics[scale=0.435]{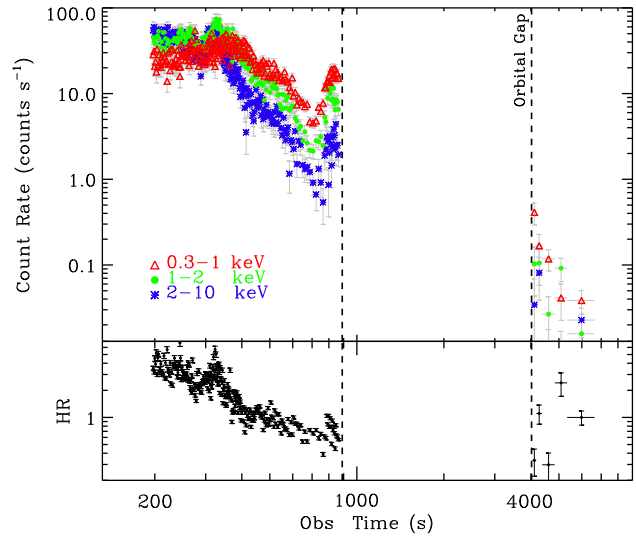}
\caption{Upper panel: steep decay portion of GRB\,081028
X-ray afterglow. The XRT signal has been split into 
3 energy bands so that the different temporal 
behaviour can be fully appreciated. Lower panel:
$(0.3-1\, \rm{keV})/(1-10\, \rm{keV})$ hardness ratio evolution with 
time. The signal clearly softens with time.
In both panels, the vertical black
dashed lines mark the orbital data gap.}
\label{Fig:xrtsteepdecaybands}
\end{figure}

\begin{figure}
\centering
\includegraphics[scale=0.526]{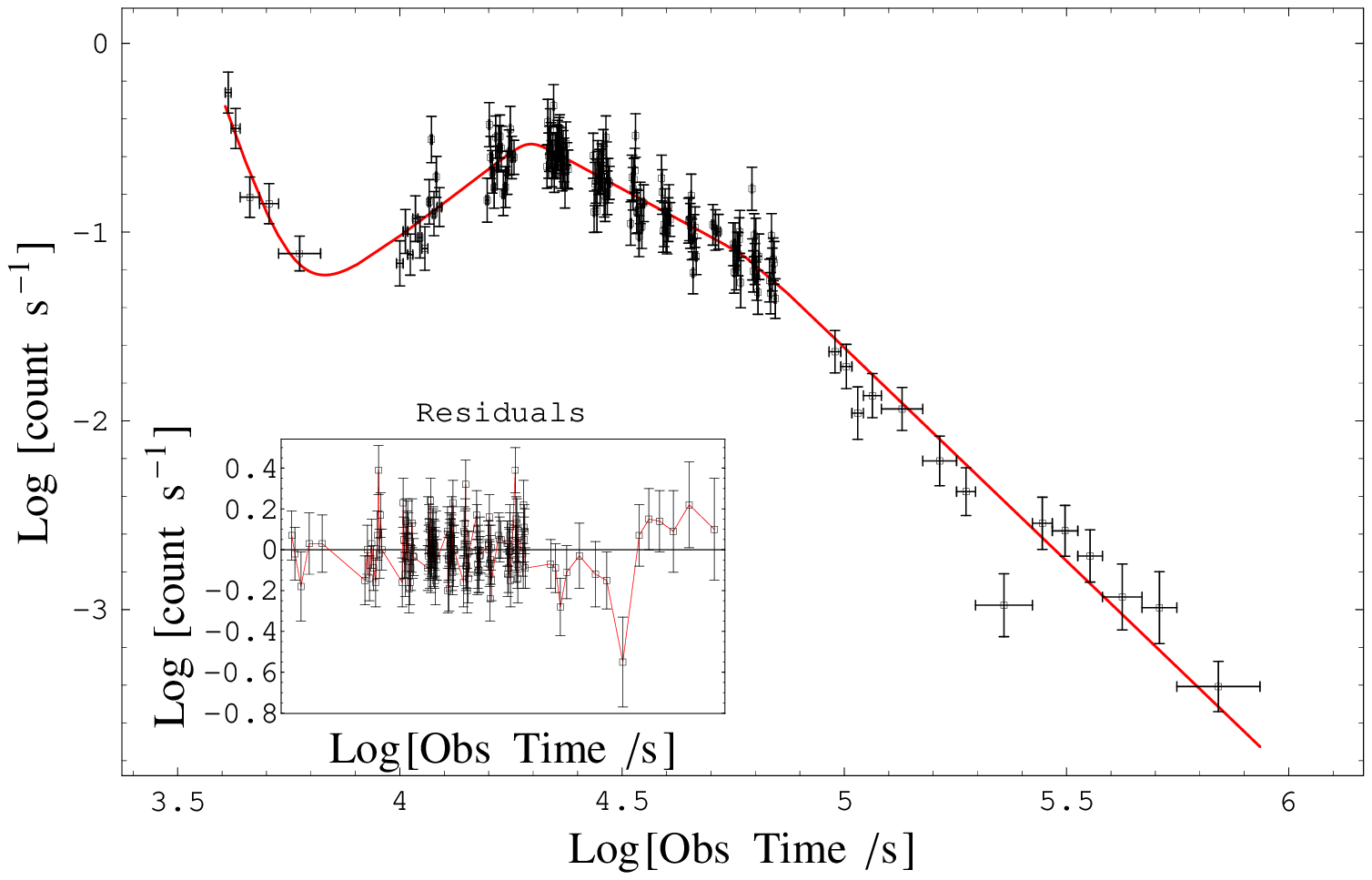}
\caption{XRT 0.3-10 keV count-rate light-curve of GRB\,081028
starting from 3 ks with best fit model superimposed (Eq. \ref{Eq:rebfin},
Table \ref{Tab:XRTrebfit}). 
\emph{Inset:} residuals with respect to the best-fit model.}
\label{Fig:XRTfitreb}
\end{figure}

The XRT (0.3-10 keV) light-curve consists of two parts: a steep decay phase
with flares and variability superimposed ($100\,\rm{s}<t<7000 \,\rm{s}$),
followed by a remarkable re-brightening with smoothly rising and 
decaying emission between 7 ks and $1000\,\rm{ks}$. The two light-curve
phases are studied separately.

GRB\,081028 is one of the rare cases in which the XRT caught the
prompt emission. The light-curve shows a flat phase up to 
$t\sim 300\,\rm{s}$ followed by a steep decay. Starting from 
$\sim690\,\rm{s}$ the light-curve is dominated by a flare which
peaks at 800 s but whose decaying phase is temporally coincident with 
the orbital data gap. The steep decay behaviour before the flare is 
inconsistent with the back-extrapolation of the post orbital 
data gap power-law decay, as shown in Fig. \ref{Fig:XRTcomponents}.
The strong spectral evolution detected by the XRT (Sect. \ref{SubSec:specXRT}) 
requires a time resolved flux calibration of the light-curve before the 
light-curve fitting procedure. In the time interval 
$320\,\rm{s}<t<685\,\rm{s}$ the 0.3-10 keV light-curve best fit is given by a 
simple power-law with $\alpha=3.6\pm0.1$ ($\chi^{2}/\rm{dof}=768.3/736$).
Figure \ref{Fig:xrtsteepdecaybands} shows the different temporal behaviour of the
detected signal when split into different energy bands: harder photons decay faster.
The 0.3-1 keV light-curve
decays following a power-law with index $\alpha \approx -2.5$; the decay steepens
to $\alpha\approx -3.5$ and $\alpha \approx -3.8$ for the 1-2 keV and 
2-10 keV signal, respectively.

During the re-brightening there is no evidence for spectral evolution in the XRT energy band
(see Sect. \ref{SubSec:specXRT}). For this reason we model the count-rate
light-curve instead of the flux calibrated one: this gives the possibility
to obtain a fully representative set of best fit parameters\footnote
{This is in general not true in cases of strong spectral evolution as 
shown in the first part of this section.} determined with the
highest level of precision. The count-to-flux calibration introduces
additional uncertainty inherited by the spectral fitting
procedure. Starting from 3 ks (the inclusion of the last part of the steep decay
is necessary to model the rising part of the re-brightening), 
the count-rate light-curve can be modelled by a power-law plus Beuermann function 
\citep{Beuermann99} where the smoothing parameter $d_1$ is left free to vary:
\begin{equation}
\rm{n_{2}}\cdot t^{\rm{c}}+n_1 \Big[\Big(\frac{t}{t_{\rm{br}_1}}\Big)^{\frac{a}{d_1}}+\Big(\frac{t}{t_{\rm{br}_1}}\Big)^{\frac{b}{d_1}}\Big]^{-d_1} 
\label{Eq:plBeuermannfree}
\end{equation}
The best fit parameters are reported in Table \ref{Tab:XRTrebfit}.
The drawback of this model is that the best-fit slopes are 
asymptotic values and do not represent the actual power-law slopes.
While due to the smooth transition between the rising and decaying phases,
this makes the comparison between observations and model predictions difficult.
Freezing $d_1$ at 0.1 to have a sharp transition results in an
unacceptable fit (P-value$\sim10^{-4}$) and suggests a light-curve steepening
around 50 ks. The possibility of a break is investigated as follows: we select 
data points starting from 20 ks and fit the data using a SPL
or a broken power-law (BPL) model. Given that the SPL and BPL models are
nested models and the possible values of the second model do not lie on the
boundary of definition of the first model parameters \citep{Protassov02}, we can 
apply an F-test: with a probability of chance improvement $\sim1\%$, we find moderate
statistical evidence for a break in the light-curve at 62 ks.
The final fitting function is given by Eq. \ref{Eq:rebfin}:
\begin{equation}\label{Eq:second}
\Bigg\{\begin{array}{ll}
{\rm{n_2}}\cdot \rm{t}^{\rm{c}}+n_1\Big[\Big(\rm{\frac{t}{t_{\rm{br1}}}}\Big)^{\frac{a}{d_1}}+\Big(\frac{t}{t_{\rm{br}_1}}\Big)^{\frac{b}{d_1}}\Big]^{-d_1}&t<40\,\rm{ks}\\
f\cdot \rm{n_3}\Big[\Big(\rm{\frac{t}{t_{\rm{br_2}}}}\Big)^{\frac{b}{d_2}}+\Big(\frac{t}{t_{\rm{br_2}}}\Big)^{\frac{e}{d_2}}\Big]^{-d_2}&t>40\,\rm{ks}\\
\end{array}
\label{Eq:rebfin}
\end{equation}
where $f$ is function of the other fitting variables and assures the continuity
of the fitting function at 40 ks. 
The light-curve of GRB\,081028 fits in this case with $\chi^{2}/\rm{dof}=147.1/143$
and an P-value=$39\%$:
the best fit parameters are reported in Table \ref{Tab:XRTrebfit} while
a plot of the result is provided in  Fig. \ref{Fig:XRTfitreb}.
The fit of the flux-calibrated light-curve gives completely consistent results.
The model predicts 
$F_{\rm{X,p}}=(1.53\pm0.08)\times 10^{-11}\,\rm{erg\,cm^{-2}\,s^{-1}}$,
where $F_{\rm{X,p}}$ is the flux at the peak of the re-brightening.

\subsubsection{Count-rate drop around 250 ks}
\label{subsubsec:drop}
The drop of the count-rate around 250 ks is worth attention: the statistical 
significance of this drop is discussed below. We select data with $t>60$ ks.
These data can be fit by a simple power-law with index 
$\alpha=1.9\pm0.2$ ($\chi^2/\rm{dof}=11.0/12$, P-value=53\%). According to this model the 
drop is not statistically significant (single trial significance of $\sim2.6~\sigma$). However,
this model under-predicts the observed rate for $t<60\,\rm{ks}$: an abrupt
drop of the count-rate during the orbital gap at 80 ks would be required in this
case. Alternatively there is not any kind of switch-off  of the
source during the orbital gap and the flux around 80 ks joins smoothly
to the flux component at $t<60$ ks, as portrayed in Fig. \ref{Fig:XRTfitreb}.
A careful inspection of the figure reveals the presence of a non random
distribution of the residuals of the last 14 points, with the points before
250 ks being systematically low and those after 250 ks being systematically
high. While this fit is
completely acceptable from the $\chi^2$ point of view, a runs test shows
that the chance probability of this configuration of residuals is less than
$0.1\%$. This would call for the introduction of a new component to model
the partial switch-off and re-brightening of the source around 250 ks. 
A possible description of the light-curve behaviour for $t>20\,\rm{ks}$
(peak time of the main re-brightening) is represented by a Beuermann plus
Beuermann function with smoothing parameters frozen to give sharp
transitions; the first decaying power-law index is frozen to $b=1.3\pm1.3$ 
while the break time of the first Beuermann component is frozen to $t_{\rm{br_2}}=62\,\rm{ks}$
as reported in Table \ref{Tab:XRTrebfit}. The light-curve decays with 
$\alpha_2=3.1\pm0.2$ ($\alpha_3=1.5\pm0.7$) for $60\,\rm{ks}<t<250\,\rm{ks}$
($t>316\,\rm{ks}$), see Fig. \ref{Fig:XRTcomponents}. This additional 
component would account for $\sim10\%$ of the total re-brightening $0.3-10$ keV
energy which is $\sim1.1\times 10^{52}\,\rm erg$.

The temporal properties of the second re-brightening seem to point to 
refreshed shocks (see e.g. \citealt{Kumar00b}; \citealt{Granot03}): the decaying 
power-laws before and after 
the drop are roughly consistent with each other but shifted upwards in the 
count-rate axis. Since at this epoch the observed X-ray frequencies are above both 
the cooling and the injection frequencies, in the standard afterglow scenario 
the X-ray flux is $\propto E_{\rm iso}^{(p+2)/4}$ independent of 
the external medium  density profile (see e.g. 
\citealt{Panaitescu00}, their appendices B and C): the observed jump in flux
would therefore require an increase of the energy in the forward shock 
by a factor of $\sim3$. Given the marginal 
statistical evidence, the properties of the second re-brightening
will not be discussed further.

\subsection{Spectral analysis of XRT (0.3-10 keV) data}
\label{SubSec:specXRT}

\begin{figure}
\centering
\includegraphics[scale=0.435]{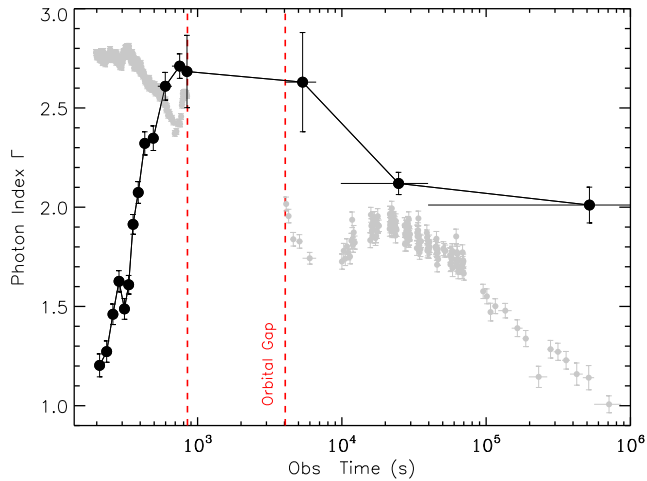}
\caption{0.3-10 keV light-curve (grey points, arbitrary units) with 
best fit 0.3-10 keV photon index superimposed (black points). Each point comes from 
the fit of a spectrum consisting of $\sim2000$
photons: the model \textsc{tbabs*ztbabs*pow} within \textsc{Xspec} with the 
intrinsic column density $N_{\rm{H,z}}$ frozen to 
$0.52\times10^{22}\,\rm{cm^{-2}}$ has been used. An exception is 
represented by the first data point after the orbital gap: see 
Sect. \ref{SubSec:specXRT} for details.
The vertical red dashed lines mark the time interval of the first
orbital gap. An abrupt change of the spectral properties
of the source temporally coincident with the onset of 
the re-brightening is apparent.  }
\label{Fig:PhotonIndex_nhtot}
\end{figure}

The very good statistics characterising the X-ray afterglow of GRB\,081028 
gives us the possibility to perform a temporally resolved spectral analysis. Figure \ref{Fig:PhotonIndex_nhtot}
shows the dramatic evolution of the photo-electrically absorbed 
simple power-law photon index with time during the first 1000 s of
observation, with $\Gamma$ evolving from $1.2$ to $2.7$. The intrinsic neutral 
Hydrogen column density $N_{\rm{H,z}}$ has been frozen to 
$0.52\times10^{22}\,\rm{cm^{-2}}$ for the reasons explained below. 
If left free to vary, this parameter shows an unphysical rising and 
decaying behaviour between 200 s and 600 s. 

The temporal behaviour of the light-curve in the time interval $4-7.5$ ks 
(after the orbital gap, see Fig. \ref{Fig:XRTcomponents}) physically connects these data 
points with the steep decay phase. We test this link from the spectroscopic point 
of view. The 0.3-10 keV spectrum extracted in this time interval contains 133 
photons. Spectral channels have been grouped so as to have 5 counts per bin and then
weighted using the Churazov method \citep{Churazov96} within \textsc{Xspec}. A fit 
with a photo-electrically absorbed power-law (\textsc{tbabs*ztbabs*pow} model) 
gives $\Gamma=2.63\pm0.25$, (90\% c.l., $\chi^{2}/\rm{dof}=25.6/23$, P-value=32\%), 
confirming
that this is the tail of the steep decay detected before the orbital gap
as shown by Fig. \ref{Fig:PhotonIndex_nhtot}.

The light-curve re-brightening around $7\,\rm{ks}$ translates into an 
abrupt change of the 0.3-10 keV spectral properties (Fig. \ref{Fig:PhotonIndex_nhtot}),
with $\Gamma$ shifting from $2.7$ to $2$. The possibility of a spectral evolution
in the X-ray band during the re-brightening is investigated as follows:
we extracted three spectra in the time intervals 7-19.5 ks (spec1, rising phase);
19.5-62 ks (spec2, pre-break decaying phase); 62 ks- end of observations (spec3,
post-break decaying phase). A joint fit of these spectra with an absorbed simple
power-law model (\textsc{tbabs*ztbabs*pow} model) where the intrinsic Hydrogen 
column density is frozen to $0.52\times10^{22}\,\rm{cm^{-2}}$ (see Sect. 
\ref{SubSec:SwiftXRTdata}) and the photon index is tied to the same value, 
gives $\Gamma=2.04\pm0.06$ with $\chi^2/\rm{dof}=118.0/167$. 
Thawing the photon indices we obtain: $\Gamma_1=2.13^{+0.14}_{-0.14}$;
$\Gamma_2=2.03^{+0.07}_{-0.07}$; $\Gamma_3=2.00^{+0.13}_{-0.12}$ 
($\chi^2/\rm{dof}=115.8/165$). Uncertainties are quoted at 90\% c.l..
The comparison of the two results
implies a chance probability of improvement
of 22\%: we conclude that there is no evidence for spectral evolution during the
re-brightening in the 0.3-10 keV energy range. The same conclusion is reached 
from the study of the $(1-10\,\rm{keV})/(0.3-1\,\rm{keV})$ hardness ratio.

\subsection{Spectral energy distribution during the re-brightening:
evolution of the break frequency}
\label{subsec:SED}

\begin{figure*}
\centering
\includegraphics[scale=0.66]{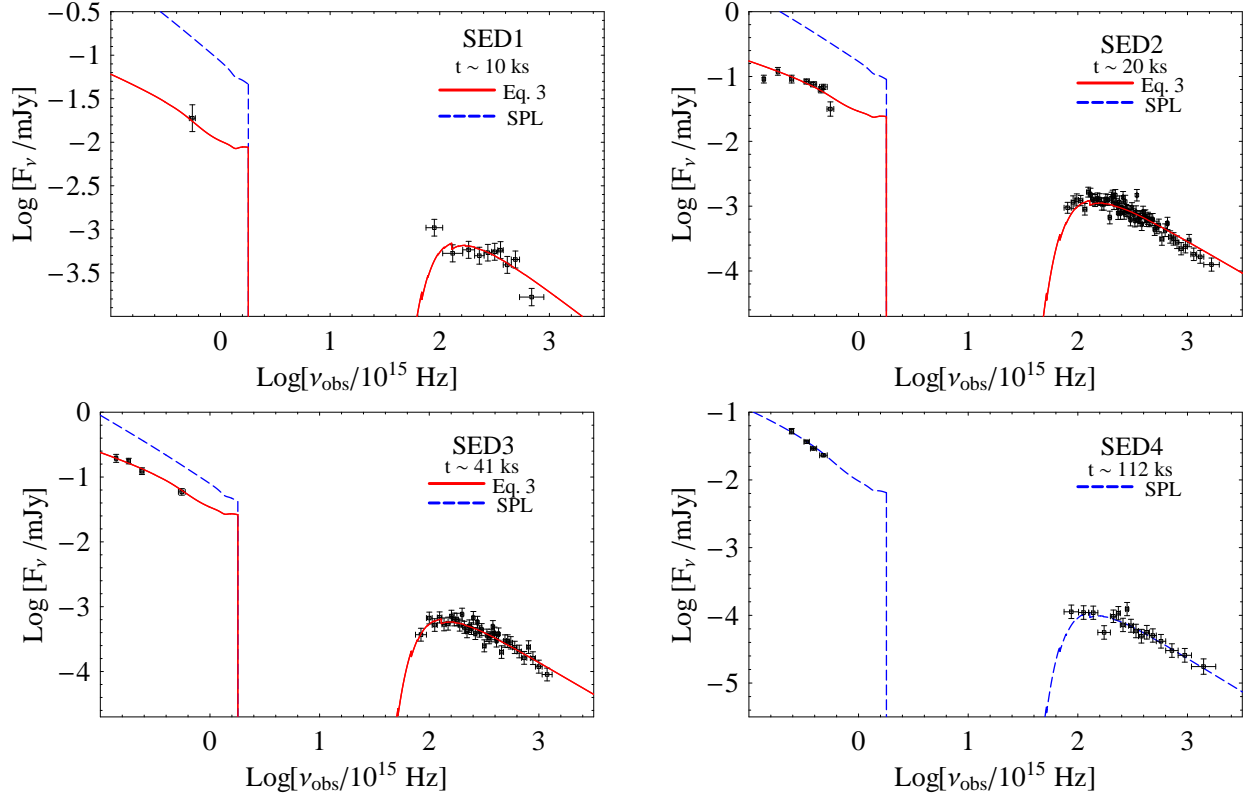}
\caption{Observer-frame SED1, SED2, SED3 and SED4 from optical to X-ray
extracted at $t\sim10\,\rm{ks}$, $t\sim20\,\rm{ks}$, $t\sim41\,\rm{ks}$ and
$t\sim112\,\rm{ks}$, respectively. Red solid line: photo-electrically absorbed 
model corresponding to Eq. \ref{Eq:specbreak}. This proved
to be the best fit model for SED1, SED2 and SED3. Blue dashed line:
photo-electrically absorbed simple power law. This is the best 
fit model for SED4. For all SEDs an SMC extinction curve at the redshift of the source 
is assumed. The best fit parameters are reported in Table \ref{Tab:SED}.}
\label{Fig:SED}
\end{figure*}

The re-brightening properties can be constrained through the study of the
temporal evolution of the spectral energy distribution (SED) from the optical
to the X-ray. We extract 4 SEDs, from the time intervals indicated by the shaded bands in Fig.
\ref{Fig:plottot_lc}: 
\begin{enumerate}
  \item SED 1 at $t\sim 10\,\rm{ks}$ corresponds to the rising portion of the
	X-ray re-brightening and includes XRT and UVOT observations;
  \item SED 2 is extracted at $t\sim20\,\rm{ks}$, peak of the X-ray re-brightening.
	It includes XRT, UVOT, GROND and NOT observations;
  \item SED 3 at $t\sim 41\,\rm{ks}$ describes the afterglow spectral energy 
	distribution during the decaying phase of the re-brightening, before the detected 
	light-curve break. It includes X-ray data from $\sim30$ ks to $\sim62$ ks,
	UVOT and PAIRITEL observations;
  \item SED 4 corresponds to the post-break decaying portion of the re-brightening,
  at $t\sim112$ ks and includes XRT and GROND observations. 
  \end{enumerate}

When necessary, optical data have been interpolated to the time of extraction 
of the SED. Uncertainties have been propagated accordingly.

At a redshift of 3.038, we expect some contamination in the spectrum from absorption 
systems located between the Earth and GRB\,081028 \citep{Madau95}. This means that
the $g'$ filter of GROND and all UVOT filters but the v band  are marginally or
strongly affected by Lyman absorption: these filters are consequently excluded from the 
following analysis. 

The Galactic and intrinsic absorption at wavelengths shorter than the 
Lyman edge are modelled using the photo-electric cross-sections of 
\cite{Morrison83}. We adopt the analytical description of the Galactic 
extinction by \cite{Pei92}, while the host galaxy absorption is assumed to 
be modelled by a Small Magellanic Cloud-like law (from \citealt{Pei92}).

An absorbed SPL model from the optical to the X-ray range is not able
to account for SED 1, SED 2 and SED 3 (Fig. \ref{Fig:SED}), while it 
gave the best fit model for SED 4.
For the first three SEDs a satisfactory fit is given by a 
broken power-law with X-ray spectral index $\beta_x\sim1$; optical spectral index
$\beta_o\sim0.5$ and  $N_{\rm{H,z}}$ consistent with the value reported in
Sect. \ref{SubSec:SwiftXRTdata} ($0.52\times10^{22}\,\rm{cm^{-2}}$). 
The best fit break frequency is found to evolve with time to lower
values following a power-law evolution with  index $\alpha\sim2$. This
evolution is faster than expected for the cooling frequency of a 
synchrotron spectrum (see e.g. \citealt{Sari98}; \citealt{Granot02}): in the following,
we identify the break frequency with the injection frequency. We freeze the
Galactic contribution to give $E(B-V)=0.03$ \citep{Schlegel98}, while leaving
the intrinsic component free to vary.

The broken power-law model has been then refined as follows. \cite{Granot02}
showed that under the assumption of synchrotron 
emission from a relativistic blast wave that accelerates the electrons 
to a power law distribution of energies $N(\gamma_e)\propto \gamma_e^{-p}$, 
it is possible to derive a  physically
motivated shape of spectral breaks. Interpreting the break frequency
as the injection frequency in the fast cooling regime, the broken 
power-law model reads (see \citealt{Granot02}, their Eq. 1):
\begin{equation}
F_{\nu}= F_{\rm{n}}\Big[\Big(\frac{\nu}{\nu_{b}}\Big)^{-s\beta_1} + \Big(\frac{\nu}{\nu_{b}}\Big)^{-s\beta_2}\Big]^{-1/s}
\label{Eq:specbreak}
\end{equation}
where  $\nu_{\rm{b}}$ and $F_{\rm{n}}$ are the break frequency 
and the normalisation, respectively;
$\beta_1= -0.5$ and $\beta_2=-p/2$ are the asymptotic spectral 
indices below and above the break under the conditions above;
$s\equiv s(p)$ is the smoothing parameter: in particular,
for an interstellar (wind) medium $s=3.34-0.82p$ ($s=3.68-0.89p$)
(\citealt{Granot02}, their Table 2). The free parameters of the final 
model are the following:
normalization of the spectrum $F_{\rm{n}}$, break frequency 
$\nu_{\rm{b}}$, power-law index of the electron distribution $p$, 
intrinsic neutral Hydrogen column density $N_{\rm{H,z}}$, 
and host reddening. The ISM or wind environments give perfectly consistent results.
We choose to quote only ISM results for the sake of brevity. For SED 4
we use an absorbed simple power-law with spectral index $-p/2$. 
The four SEDs are first fit separately; as a second step we perform a joint fit
where only the spectral normalisation and break frequency are  
free to take different values in different spectra. We find fully 
consistent results with improved uncertainties thanks to the tighter 
constraints imposed by the joint fit. The best fit results are
reported in Table \ref{Tab:SED} and portrayed in Fig. \ref{Fig:SED}.\\

\begin{table}\footnotesize
\begin{center}
\begin{tabular}{ccc}
\hline
SED & Parameter & Value\\
\hline
1,2,3,4 &p &  $1.97\pm0.03$\\
1&$\rm{Log}_{10}(\nu_{\rm{b}}/10^{15}\rm{Hz})$  &$2.0\pm0.1$\\
2&$\rm{Log}_{10}(\nu_{\rm{b}}/10^{15}\rm{Hz})$  &$1.4\pm0.1$\\
3&$\rm{Log}_{10}(\nu_{\rm{b}}/10^{15}\rm{Hz})$  &$0.4\pm0.1$\\
\hline
$\chi^{2}/\rm{dof}$&\multicolumn{2}{c}{134.7/138}\\
P-value & \multicolumn{2}{c}{56\%}\\
\hline
\end{tabular}
\caption{Best fit parameters for the simultaneous fit of SED1, SED2,
SED3 and SED4. For SED1, SED2 and SED3 the emission model is expressed
by Eq. \ref{Eq:specbreak}, while for SED4 we used a simple power-law with 
spectral index $p/2$. The spectral normalisations and  break 
frequencies have been left free to take different values in different spectra.
The intrinsic neutral Hydrogen column value is found to be consistent with the
value inferred from the X-ray spectra.}
\label{Tab:SED}
\end{center}
\end{table}

\begin{figure}
\centering
\includegraphics[scale=0.6]{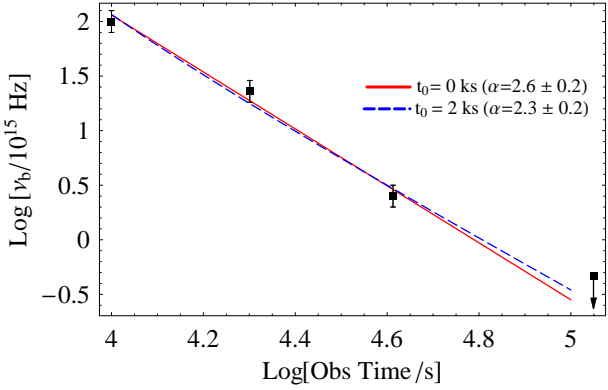}
\caption{Spectral break frequency (See Eq. \ref{Eq:specbreak}) evolution with time
as found from a simultaneous fit of SED1, SED2, SED3 and SED4 with 
best fit models superimposed. Red solid line (blue dashed line): simple power-law
with zero time $t_0$=0 ks (2 ks) and best power-law index $\alpha=2.6$ (2.3). 
The satisfactory fit of SED4 with a simple power-law provides the upper limit shown.}
\label{Fig:BreakFrequency}
\end{figure}

The spectral break frequency $\nu_{\rm{b}}$ evolves with time to lower
values, as shown in Fig. \ref{Fig:BreakFrequency}. The 
consistency of SED4 optical and X-ray data with a simple power-law model with 
index $-p/2$, 
suggests that the break frequency has crossed the optical band by the 
time of extraction of SED4. This translates into 
$\rm{Log}_{10}(\nu_{\rm{b}}/10^{15}\rm{Hz})<-0.33$ for $t>112\,\rm{ks}$. 
The decrease of the break frequency with time can be modelled by a simple
power-law function: this leads to an acceptable fit 
($\chi^2/\rm{dof}=1.4/1$, P-value=24\%)
with best fit index $\alpha=2.6\pm0.2$. Using $t_{0}=2$ ks
as zero time of the power-law model we obtain: $\alpha=2.3\pm0.2$
($\chi^2/\rm{dof}=2.1/1$, P-value=15\%).

The fit implies a limited rest frame optical extinction
which turns out to be $E(B-V)_{\rm{z}}\sim 0.03$.
A $3-\sigma$ upper limit can be derived from the joint fit of
the four SEDs, leaving all the parameters but the one related to the optical
extinction free to vary. The upper limit is computed as the value which increases
the $\chi^2$ by a $\Delta\chi^2$ corresponding to a $3~\sigma$ c.l..
This procedure leads to: $A_{V,z}<0.22$.

\subsection{Peak energy evolution with time}
\label{SubSec:Epeak}

\begin{figure}
\centering
    \includegraphics[scale=0.58]{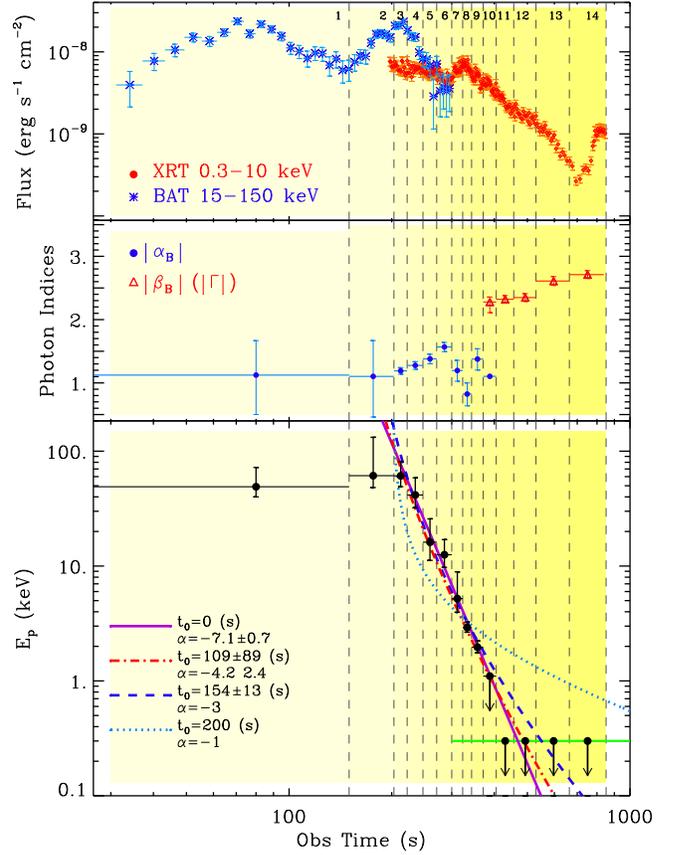}
      \caption{Time resolved combined analysis of XRT and BAT data.
Upper panel: BAT 15-150 keV and XRT 0.3-10 keV flux light-curves. No extrapolation
of the BAT data into the XRT energy range has been done. The vertical dashed lines
mark the intervals of extraction of the spectra: these are numbered
according to Table \ref{Tab:Epeak}, first column. Central panel: best fit photon
indices evolution with time. Lower panel: best fit $E_{\rm{p}}$ parameter as a 
function of time. The decay has been fit with a simple power-law model starting
from 200 s from trigger: $E_{\rm{p}}(t)\propto (t-t_{0})^{\alpha}$.
Starting from 405 s $E_{\rm{p}}$ is likely to be outside the XRT energy range: 
$E_{\rm{p}}<0.3\,\rm{keV}$ (solid green line).
 }
\label{Fig:Epeak}
\end{figure}

\begin{table*}\footnotesize
\begin{center}
\begin{tabular}{ccccccccccc}
\hline
Interval&$t_i$ & $t_f$ &        & Model & $\alpha_{\rm{B}}$& $\beta_{\rm{B}}(\Gamma)$&$E_{\rm{p}}$&$\chi^2/\rm{dof}$ & P-value\\
&(s)   & (s)   &        &       &         &        &  (keV)    &                 &         \\
\hline
3&203   &	222   & BAT+XRT& Cutpl & $1.19\pm0.05$& ---& $61.0^{+20.0}_{-11.9}$&100.4/112 & 77\%\\ 
 &      &             &        &  Pl   & ---&$1.37\pm0.02$& ---& 147.3/114& 2\%\\          
\hline
4&222   & 247   & BAT+XRT& Cutpl & $1.28\pm0.06$& ---& $41.5^{+17.1}_{-9.4}$&80.5/101& 93\%\\ 
 &      &             &        &  Pl   & ---& $1.44\pm0.03$& ---& 108.2/102& 31\%$^{+}$\\
\hline
5&247   & 271   & BAT+XRT& Cutpl & $1.38\pm0.17$& ---& $16.1^{+9.6}_{-4.9}$ &77.4/88& 78\%\\ 
 &      &             &        &  Pl   & ---& $1.54\pm0.04$& ---& 96.5/89& 3\%$^{+}$\\
\hline
6&271   & 300   & BAT+XRT& Cutpl & $1.57\pm0.07$& ---& $12.5^{+4.5}_{-2.7}$ &129.8/91& 1\%\\
 &      &             &        &  Pl   & ---& $1.76\pm0.03$& ---& 158.6/92& 0.001\%\\
\hline
7&300   & 323   & XRT    & Cutpl & $1.20\pm0.16$& ---& $5.2^{+3.7}_{-1.3}$  &77.1/81& 60\%\\
 &      &             &        &  Pl   & ---& $1.49\pm0.05$& ---& 87.6/82& 32\%\\
\hline
8&323   & 343   & XRT    & Cutpl & $0.82\pm0.18$& ---& $2.9^{+0.3}_{-0.3}$& 78.7/83& 61\%\\
 &      &             &        &  Pl   & ---&$1.61\pm0.05$& ---& 149.8/84& 0.001\%\\
\hline
9&343   & 371   & XRT    & Cutpl & $1.38\pm0.17$& ---& $2.0^{+0.3}_{-0.3}$& 94.3/84& 15\% \\
 &      &             &        &  Pl   & ---&$1.91\pm0.05$& ---& 131.2/82& 0.1\%\\
\hline
10&371   & 405   & XRT    & Band  & $\sim1.10$&$2.3^{+0.1}_{-0.2}$& $<1.1$& 82.4/77&31\%\\ 
 &      &             &        &  Cutpl   & $1.81\pm0.016$& ---&$1.0^{+0.3}_{-0.9}$&102.4/78& 3\%\\
 &      &             &        &  Pl   & ---& $2.07\pm0.06$& ---& 109.7/79& 1\%\\
\hline
11&405   & 456   & XRT    & Pl    & ---& $2.32\pm0.06$& ---& 100.1/78& 5\%\\
\hline
12&456   & 530   & XRT    & Pl    & ---& $2.34\pm0.06$& ---& 103.3/79& 3\%\\
\hline
13&530   & 664   & XRT    & Pl    & ---& $2.61\pm0.07$& ---&  98.1/76 & 5\%\\
\hline
14&664   & 838   & XRT    & Pl    & ---& $2.71\pm0.06$& ---&  89.3/73 & 7\%\\
\hline
15&838   & 851   & XRT    & Pl    & ---& $2.68\pm0.18$& ---&  15.7/10& 1\%\\
\hline
\end{tabular}
\caption{Best fit parameters derived from the spectral modelling of XRT and BAT
data using photo-electrically absorbed models (\textsc{tbabs*ztbabs} within 
\textsc{Xspec}). The BAT and XRT normalisations are always tied to the same
value. Three different models have been used: a simple power-law (Pl); 
a cut-off power law and a Band function  both with the peak energy of the $\nu F_{\nu}$ 
spectrum as free parameter. From left to right: name of the interval
of the extraction of the spectrum we refer to throughout the paper (intervals 1 and 2
correspond to Pulse 1 and Pulse 2 of Table \ref{Tab:BATspec}); 
start and stop time of extraction of each spectrum; energy range of the fit: ``XRT+BAT"
stands for a joint BAT-XRT data fitting; model used; best fit low and high energy 
photon indices for a Band or Cutpl 
power-law or best fit photon index $\Gamma$ for a Pl model; statistical information 
about the fit. The $^{+}$ symbol indicates an apparent trend in the residuals of the fit.}
\label{Tab:Epeak}
\end{center}
\end{table*}

The consistency of the prompt BAT spectrum with a cut-off power-law 
(Sect. \ref{Tab:BATspec}) and the spectral variability detected in the 
XRT energy range (Sect. \ref{SubSec:specXRT}, Fig. \ref{Fig:PhotonIndex_nhtot}) 
suggests that the peak of the $\nu F_{\nu}$ spectrum is moving through the 
BAT+XRT bandpass. To follow the spectral evolution, we time slice the BAT and XRT
data into 14 bins covering the 10-851 s time interval. The spectra are then fit
within \textsc{Xspec} using a Band function (\textsc{ngrbep}) or a cut-off 
power-law (\textsc{cutplep}) with $E_{\rm{p}}$ as free parameter; alternatively
a simple power law is used. Each model is absorbed by a Galactic 
(hydrogen column density frozen to $3.96\times10^{20}\,\rm{cm^{-2}}$) and 
intrinsic component ($N_{\rm{H,z}}$ frozen to $0.52\times10^{22}\,\rm{cm^{-2}}$,
see Sect. \ref{SubSec:specXRT}).
When possible we take advantage of the simultaneous BAT and XRT
observations, performing a joint BAT-XRT spectral fit. The normalisation for 
each instrument is always tied to the same value.
The best fit parameters are reported in Table \ref{Tab:Epeak}:
the simple power law model gives a poor description of the spectra up to 
$\sim400$ s, as the curvature of the spectra requires a cut-off power-law
or a Band function. In particular, this is the case when
the high energy slope enters the XRT bandpass. 
The $E_{\rm{p}}$ parameter is well constrained and evolves to lower energies
with time; at the same time both the high and low energy photon indices
are observed to gradually vary, softening with time (Fig. \ref{Fig:Epeak}).
The $E_{\rm{p}}$ decay with time can be modelled by a simple
power-law starting $\sim200$ s after trigger: $E_{\rm{p}}\propto(t-t_0)^\alpha$.
The best fit parameters are reported in Table \ref{Tab:Epeakfit}. 

The uncertainty of the inter-calibration of the BAT and XRT has 
been investigated as possible source of the detected spectral evolution
as follows. For each time slice, we multiply the fit model by a 
constant factor which is frozen to 1 for the BAT data. For XRT, this factor
is left free to vary between 0.9 and 1.1, conservatively allowing the XRT 
calibration to agree within $10\%$ with the BAT calibration. The best fit 
parameters found in this way are completely consistent with the ones listed 
in Table \ref{Tab:Epeak}.
The inter-calibration is therefore unlikely to be the main source of the 
observed evolution.

\begin{table}\footnotesize
\begin{center}
\begin{tabular}{cccc}
\hline
 $t_0$ &$\alpha$   & $\chi^2/\rm{dof}$ &Model \\
\hline
 0 (s)& $-7.1\pm0.7$&1.7/4&---\\
 $109\pm89$ (s)&$-4.2\pm2.4$&2.5/5&---\\
 $154\pm13$ (s)&$-3$ &3.2/4& Adiabatic cooling\\
 200 (s)& $-1$ &42.1/4 &High latitude emission\\
\hline
\end{tabular}
\caption{Best fit parameters and statistical information 
for a simple power-law fit to the $E_{\rm{p}}$
decay with time starting from 200 s after trigger: 
$E_{\rm{p}}\propto(t-t_0)^\alpha$.}
\label{Tab:Epeakfit}
\end{center}
\end{table}

\section{Discussion}
\label{sec:discussion}
In GRB\,081028 we have the unique opportunity to observe a smoothly 
rising X-ray afterglow after the steep decay: this is the
first (and unique up to July 2009) long GRB \emph{Swift}-XRT light-curve 
where a rise with completely different properties to typical X-ray flares
(\citealt{Chincarini07}, \citealt{Falcone07}) is seen at $t\geq 10\,\rm{ks}$.
At this epoch,
canonical X-ray light-curves (e.g., \citealt{Nousek06}) typically show a 
shallow decay behaviour with flares superimposed in a few cases
(Chincarini et al., in prep.): only in GRB\,051016B is a 
rising feature detected at the end of the steep decay \footnote
{See the \emph{Swift}-XRT light-curve repository, \cite{Evans09} and
\cite{Evans07}.}.
In this case,
the sparseness of the data prevents us from drawing firm conclusions, so that a flare
origin of the re-brightening cannot be excluded.

The very good statistics of GRB\,081028 allows us to track the detailed
spectral evolution from $\gamma$-rays to X-rays, from the
prompt to the steep decay phase: this analysis fully qualifies the 
steep decay as the tail of the prompt emission. At the same time,
it reveals that the steep decay and the following X-ray re-brightening
have completely different spectroscopic properties (Fig.
\ref{Fig:PhotonIndex_nhtot}): this, together with the temporal
behaviour, strongly suggests that we actually see two different emission 
components overlapping for a small time interval, as was first suggested 
by \cite{Nousek06}.

The small overlap in time of the two components is the key 
ingredient that observationally allows the detection of the rising phase:
this can be produced by either a steeper than usual steep decay or a 
delayed onset of the second component. We tested both possibilities
comparing GRB\,081028 properties against a sample of 32 XRT light-curves
of GRBs with known redshift and for which the steep-flat-normal decay
transitions can be easily identified. While 63\% of the GRBs are
steeper than GRB\,081028 ($\alpha_1\sim2$), no GRB in the sample shows a 
rest frame steep-to-flat transition time greater than $1\,\rm{ks}$,
confirming in this way the ``delayed-second-component" scenario.
Alternatively, the peculiarity of GRB\,081028 could reside in a steeper
than usual rise of the second component: unfortunately this possibility 
cannot be tested. 

This section is organised as follows: in Sect. \ref{sebsec:specevolsteep}
we discuss the spectral evolution during the prompt and steep decay phases
in the context of different interpretations. The afterglow modelling of Sect. 
\ref{sec:aftmodellingtot} favours an 
off-axis geometry: however, this seems to suggest a different physical origin of the prompt 
plus steep decay and late re-brightening components. This topic is further investigated from
the prompt efficiency perspective in Sect.  \ref{subsec_prompt_eff}.
 
\subsection{Spectral evolution during the prompt and steep decay emission}
\label{sebsec:specevolsteep}
\begin{figure}
\centering
    \includegraphics[scale=0.44]{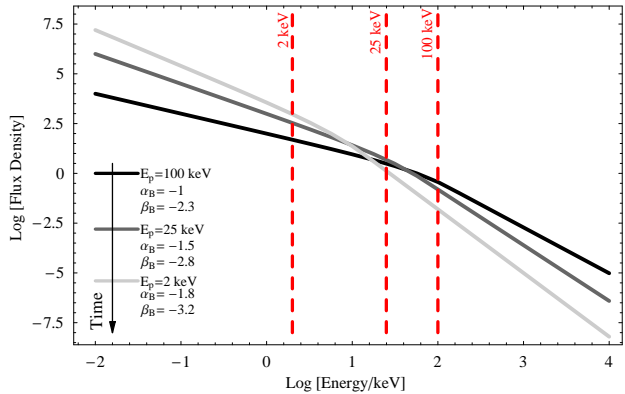}
      \caption{Qualitative description of the spectral evolution
      with time detected in GRB\,081028 from the prompt to the steep decay
phase: the peak energy ($E_{\rm p}$) moves to lower energies while 
both the high and low energy components soften with time. Arbitrary flux density
units are used.}
\label{Fig:BandEvolution}
\end{figure}

The evolution of the peak energy $E_{\rm p}$ of the $\nu F_{\nu}$ spectrum
from the $\gamma$-ray to the X-ray band described in Sect. \ref{SubSec:Epeak} 
offers the opportunity to constrain the mechanism responsible 
for the steep decay emission. 

Spectral evolution through the prompt and steep decay phase has been
noted previously, with the $E_{\rm p}$ tracking both
the overall burst behaviour and individual prompt pulse structures
(see e.g., \citealt{Peng09} for a recent time resolved
spectral analysis of prompt pulses). In particular, 
\cite{Yonetoku08} find $E_{\rm p}\propto t^{\sim-3}$ for GRB\,060904A; 
\cite{Mangano07} model the prompt to steep decay transition of GRB\,060614 
with a Band (or cut-off power-law) spectral model with $E_{\rm p}$ evolving 
as $t^{\sim-2}$, while \cite{Godet07} and \cite{Goad07b} report on the 
evolution of the $E_{\rm p}$ through the XRT energy band during single X-ray
flares in GRB\,050822 and GRB\,051117, respectively. A decaying $E_{\rm p}$
was also observed during the 0.3-10 keV emission of GRB\,070616 
\citep{Starling08}.

The detection of strong spectral 
evolution violates the prediction of the curvature effect in its simplest
formulation as found by \cite{Zhang07} in 75\% of the analysed GRBs tails: 
this model assumes the instantaneous spectrum at the 
end of the prompt emission to be a simple power-law of spectral index
$\beta$ and predicts the $\alpha=2+\beta$ relation, where $\beta$ is not supposed to vary
(see e.g., \citealt{Fenimore96}; \citealt{Kumar00}). 
The curvature effect of a comoving Band spectrum predicts instead $E_{\rm p}\propto
t^{-1}$ and a time dependent $\alpha=2+\beta$ relation (see e.g., 
\citealt{Genet09}; \citealt{Zhang09}): from Fig. \ref{Fig:Epeak},
lower panel and Table \ref{Tab:Epeakfit} it is apparent that the observed 
$E_{\rm p}\propto t^{-7.1\pm0.7}$
is inconsistent with the predicted behaviour even when we force the
zero time of the power-law fit model to be $t_0=200\,\rm s$,
peak time of the last pulse detected in the 15-150 keV energy range, 
as prescribed by \cite{Liang06}. However, a more realistic version of the HLE 
might still fit the data: a detailed 
modelling is beyond the scope of this paper and will be explored
in a future work.

The adiabatic expansion cooling of the gamma-ray producing source, 
which lies within an angle of $1/\gamma$ (where $\gamma$ is the 
Lorentz factor of the fireball) to the observer line of sight, 
has also been recently proposed as a possible mechanism responsible for 
the steep decay \citep{Duran09}. This process gives a faster 
temporal evolution of the break frequency as it passes through the X-ray 
band: typically  $E_{\rm p}\propto t^{-3}$.
Two fits to the data have been done, the first fixing 
the break evolution to $t^{-3}$ and the other one leaving $t_0$ and the break 
temporal evolution as free parameters.  Both fits are consistent with the 
adiabatic cooling expectation and set $t_0$ close to the beginning of the 
last pulse in the BAT light-curve (see Table \ref{Tab:Epeakfit}). 
However,  the adiabatic expansion cooling of a thin ejecta predicts a light-curve decay 
that is linked to the spectral index $\beta$ by the relation $\alpha=3\beta+3$,
where $\alpha$ is the index of the power-law decay. Since $\alpha_{\rm{obs}}
\sim 3.6$ this would imply $\beta\sim0.2$ which is much harder than  
observed (Sect. \ref{SubSec:specXRT}).  This makes the adiabatic cooling
explanation unlikely. 

Both the curvature effect and the adiabatic model 
assume an abrupt switch-off of the source after the end of the prompt
emission: the inconsistency  of observations with both models 
argues against this conclusion and favours models where the X-ray
steep decay emission receives an important contribution from the 
continuation of the central engine activity. In this case, the steep decay
radiation  reflects  (at least partially) the decrease in power of the GRB jet.
An interesting possibility is given by a decrease of power originating
from a decrease in the mass accretion rate \citep{Kumar08}.

Alternatively, the observed spectral softening could be caused 
by cooling of the plasma whose cooling frequency identified with $E_{\rm p}$
decreases with time as suggested by \cite{Zhang07}. 

While the spectral peak is moving, we also observe a softening of the
spectrum at frequencies both below and above the peak when our data
allow us to constrain the low and high energy slopes of a comoving Band
spectrum. A softening of the low energy index in addition to the 
$E_{\rm p}$ evolution has been already observed in the combined BAT+XRT 
analysis of GRB\,070616 (\citealt{Starling08}, their Fig. 5).
This result is consistent with the finding that while short GRBs have a 
low energy spectral component harder than long GRBs
(i.e., $|\alpha_{\rm B,short}|<|\alpha_{\rm B,long}|$, where 
$\alpha_{\rm B}$ is the low energy photon index of the \citealt{Band93} function), 
no difference is 
found in the  $\alpha_{\rm B}$ distribution of the two classes of GRBs 
when only the first 1-2 s of long GRB prompt emission is considered
\citep{Ghirlanda09}: a soft evolution of the $\alpha_{\rm B}$ parameter
with time during the $\gamma$-ray prompt emission of long GRBs is therefore required.
Our analysis extends this result to the X-ray regime and
indicates the softening of both the high and low spectral components
from the prompt to the steep decay phase. The overall spectral evolution 
is qualitatively represented in Fig. \ref{Fig:BandEvolution}. 

\subsection{Afterglow modelling}
\label{sec:aftmodellingtot}
\subsubsection{Failure of the dust scattering, reverse shock and onset of the afterglow models}
\label{subsec_aftmod}
This subsection is devoted to the analysis of the X-ray re-brightening
in the framework of a number of different theories put forward
to explain the shallow decay phase of GRB afterglows.

According to the dust scattering model \citep{Shao07} the shallow 
phase is due to prompt photons scattered by dust grains in 
the burst surroundings: this models predicts a strong spectral 
softening with time and a non-negligible amount of dust extinction
which are usually not observed \citep{Shen09}.
Both predictions are inconsistent with our data.

A spherical flow is expected 
to give rise to a peak of emission when the 
spectral peak enters the energy band of observation
(see e.g., \citealt{Granot02}): the SED analysis
of Sec. \ref{subsec:SED} clearly shows that $E_{\rm p}$ was already 
below the X-ray band during the X-ray rising phase, well before the 
peak, thus ruling out the passage of the break frequency through the
X-ray band as an explanation of the peak in the X-ray light-curve.

\cite{Sari99b} argue that the reverse shock has a much lower 
temperature and is consequently expected to radiate at lower 
frequencies than the forward shock, even if it contains an amount 
of energy comparable to the GRB itself, making a reverse shock origin 
of the X-ray re-brightening unlikely. However, following \cite{Genet07}, 
in the case of ejecta having a tail of Lorentz factor 
decreasing to low values, if a large amount of the energy dissipated 
in the shock ($\epsilon_e$ near its equipartition value) is transferred 
to only a fraction of electrons (typically $\xi_e \sim 10^{-2}$), then 
the reverse shock radiates in X-rays. In this case, it can also produce a plateau or 
re-brightening, the latter being more often obtained in a constant 
density external medium, that qualitatively agrees with the GRB\,081028 afterglow.

Alternatively, the detected light-curve peak could be the onset of the
afterglow: in this scenario, the rising (decaying) flux is to be 
interpreted as pre-deceleration (post-deceleration) forward shock 
synchrotron emission. 
The observed break frequency scaling $\nu_{\rm b} \propto t^{-2.6 
\pm 0.2}$ is inconsistent with the expected cooling frequency evolution
$\nu_{\rm c}\propto t^{-1/2}$ or $\nu_{\rm c}\propto t^{1/2}$ for an 
ISM or a wind environment, respectively (see e.g. \citealt{Granot02}).
We therefore consider a fast cooling scenario where 
$\nu_{\rm b}\equiv \nu_{\rm m}$. 
The initial afterglow signal from a thick shell is likely to overlap
in time with the prompt emission \citep{Sari99b}, so that it would
have been difficult to see the smoothly rising X-ray re-brightening
of GRB\,081028. For this reason only the onset of the forward shock 
produced by thin shells will be discussed.
Following \cite{Sari99b}, the observed peak of the X-ray
re-brightening implies a low initial fireball Lorentz factor
$\gamma_{0}\sim 75(n_0\epsilon_{\gamma,0.2})^{-1/8}$, where $n_0=n/(1\,
\rm{cm^{-3}})$ is the circumburst medium density and 
$\epsilon_{\gamma,0.2}=\epsilon_{\gamma}/0.2$ is the radiative efficiency.
Since the X-ray frequencies are always above the injection frequency $\nu_{\rm m}$, 
the X-ray light-curve should be proportional to $t^2\gamma(t)^{4+2p}$: 
during the pre-deceleration phase this means $F_X\propto t^{2}$  for 
an ISM and $F_X\propto t^{0}$ for a wind. The ISM scaling
is consistent with the observed power-law scaling $\propto t^{1.8\pm0.3}$
if a sharp transition between the rising and the decaying part of the
re-brightening is required. The asymptotic value of the power-law index during
the rising phase is instead steeper than 2, as indicated by the fit of the 
re-brightening where the smoothing parameter is left free to vary: $\propto t^{4.5\pm3.3}$ 
(see Table \ref{Tab:XRTrebfit} for details).
The injection frequency is expected to scale as $\nu_m \propto \gamma(t)^{4-k} t^{-k/2}$, 
where the density profile scales as $R^{-k}$.
This implies that for radii $R<R_{\gamma}$
(or $t<t_{\gamma}$) $\nu_{\rm m}\propto t^{0}$ for an ISM and 
$\nu_{\rm m}\propto t^{-1}$ for a wind, while for $R>R_{\gamma}$
($t>t_{\gamma}$) the fireball experiences a self-similar deceleration 
phase where $\gamma\propto t^{-3/8}$ for an ISM and $\gamma\propto t^{-1/4}$ for a wind,
 and $\nu_{\rm m}\propto t^{-3/2}$ in both cases.
$R_{\gamma}$ is the radius where a surrounding mass smaller than the 
shell rest frame mass by a factor $\gamma_{0}$ has been swept up;
$t_{\gamma}$ is the corresponding time: for GRB\,081028 $t_{\gamma}\sim 
20\,\rm ks$ (observed peak of the re-brightening). While for $t>t_
{\gamma}$ the observed evolution of the break frequency
is marginally consistent with $t^{-3/2}$, it is hard to 
reconcile the observed $\nu_{\rm m} \propto t^{-\alpha}$ with $\alpha \sim 2.6-2.4$
decay with the expected constant behaviour or $\propto t^{-1}$ decay for $t<t_{\gamma}$. 
This argument makes the interpretation of the re-brightening as onset of the 
forward shock somewhat contrived. Moreover, the identification of 
$t=20\,\rm{ks}$ with the deceleration time is also disfavoured by the 
earlier very flat optical light-curve. An alternative explanation is discussed
in the next subsection.

\subsubsection{The off-axis scenario}
\label{subsec_aftmod_offaxis}
For a simple model of a point source at an angle of
$\theta$ from the line of sight, moving at a Lorentz factor $\gamma
\gg 1$ with  $\gamma\propto R^{-m/2}$, where $R$ is its radius, the
observed time is given by:
\begin{equation}
t = \frac{R}{2c\gamma^2}\left(\frac{1}{1+m}+\gamma^2 \theta^2\right)\
\end{equation}
The peak in the light curve occurs when the beaming cone widens enough
to engulf the line of sight, $\gamma(t_{\rm{peak}}) \sim 1/\theta$, so that
before
the peak $t \approx R\theta^2/2c \propto R$. We consider an
external density that scales as $R^{-k}$ (with $k < 4$) for which $m =
3-k$. When the line of sight is outside the jet aperture, at an angle $
\theta$ {\it from the outer edge of the jet}, the
emission can be approximated to zeroth order as arising from a point
source located at an angle $\theta$ from the line of sight \citep{Granot02b}.
We have:
\begin{equation}\label{Eq:offaxis1}
\frac{t_0}{t} \sim \frac{\nu}{\nu_0}=\frac{1-\beta}{1-\beta \cos \theta}
\equiv a_{\rm aft} \approx \frac{1}{1+\gamma^2 \theta^2}
\end{equation}
where $\beta = (1-\gamma^{-2})^{1/2} = v/c$ and
the subscript $0$ indicates the $\theta = 0$ (on-axis) condition. 
The observed flux is given by
\begin{equation}\label{eq_flux_theta}
F_{\nu}(\theta, t) \approx a_{\rm aft}^3 F_{\nu/a}(0,at)\
\end{equation}
and peaks when $\gamma\sim 1/\theta$. In the following we use the notations 
$a_{\rm aft} \approx 1/(1+\gamma^2 \theta^2)$; $a$ for the particular case where 
$\gamma = \Gamma_0$ (where $\Gamma_0$ is the initial Lorentz factor of the 
fireball): $a \approx 1/(1+\Gamma_0^2 \theta^2)$.

For $t\ll t_{\rm{peak}}$, $\gamma\theta \gg 1$ and therefore $a_{\rm{aft}}
\approx (\gamma\theta)^{-2} \propto \gamma^{-2} \propto R^{3-k} \propto
t^{3-k}$. In this condition the local emission from a spherically expanding
shell and a jet would be rather similar to each other, and 
the usual scalings can be used for an on-axis viewing angle 
(e.g., \citealt{Granot02}):

\begin{equation}\label{nu_01}
\nu_{m,0} \propto R^{-3(4-k)/2} \propto t^{-3/2}
\end{equation}
\begin{equation}\label{nu_02}
\nu_{c,0} \propto R^{(3k-4)/2} \propto t^{(3k-4)/(8-2k)}
\end{equation}
with respective off-axis frequencies:
\begin{equation}\label{nu_03}
\nu_m \approx  a\,\nu_{m,0} \propto R^{(k-6)/2} \propto t^{(k-6)/2}
\end{equation}
\begin{equation}\label{nu_04}
\nu_c \approx  a\,\nu_{c,0} \propto R^{(2+k)/2} \propto  t^{(2+k)/2}
\end{equation}
For $t>t_{\rm{peak}}$,~ $a_{\rm{aft}} \approx 1$ and $\nu \approx
\nu_0$, so that the break frequencies have their familiar temporal
scaling for a spherical flow (eq.~\ref{nu_01} and \ref{nu_02})\footnote{While
these expressions are derived for a spherical flow, they are
reasonably valid even after the jet break time $t_{\rm jet}$ as
long as there is relatively very little lateral expansion as shown
by numerical simulations (see e.g., \citealt{Granot01c};
\citealt{Zhang09b} and references therein).}.

For a uniform external medium ($k = 0$), $\nu_c \propto t$ and
$t^{-1/2}$ before and after the peak, respectively, while for a
stellar wind environment ($k = 2$) the corresponding temporal scalings
are $t^2$ and $t^{1/2}$. In both cases this is inconsistent with the
observed rapid decrease in the value of the break frequency
($\nu_{\rm{b}}\propto t^{-2.6}$) unless we require a very sharp
increase in the magnetic field within the emitting region due
to a large and sharp increase in the external density \citep{Nakar07}. 
We consider this possibility unlikely (see Sect. \ref{subsec_aftmod}).

Alternatively, the break frequency could be $\nu_m$, for a fast
cooling spectrum where $\nu_c$ is both below $\nu_m$ and below the
optical. In this case, for $t<t_{\rm{peak}}$ we have $\nu_m\propto t^{-3}$
($t^{-2}$) for a $k = 0$ ($k= 2$) environment; after the peak
$\nu_m \propto t^{-3/2}$ independent of $k$. Since we observe
$\nu_{\rm b} \propto t^{-2.6 \pm 0.2}$ (or $\nu_{\rm b}\propto
(t-t_0)^{-2.3 \pm 0.1}$ with $t_0=2\,\rm ks$)
over about a decade in time around the light-curve peak, this is
consistent with the expectations for a reasonable value of $k$.

Constraints on the model parameters are derived as follows: given that
we see only one break frequency in our SEDs, which we identify with $\nu_{\rm
m}$, we must require $\nu_{\rm c}<\nu_{\rm opt}(\approx10^{15}\,\rm Hz)$.
The tightest constraints are derived at $t_{\rm peak}$, when $\nu_{\rm c}$
reaches its maximum value (it increases with time before
$t_{\rm peak}$ and decreases with time after $t_{\rm peak}$ for 
$k<4/3$). From
\cite{Granot02},
their Table 2, spectral break 11, this means:
\begin{equation} \label{eq_cdition_nuc<nuopt}
\epsilon_B^{3/2} n_0 E_{k,54}^{1/2}(1+Y)^2 > 10^{-3}\
\end{equation}
where $\epsilon_{\rm{B}}$ is the fraction of the downstream  (within the shocked region)
internal energy going into the magnetic field; $n_0 = n/(1\;{\rm cm^{-3}})$
is the
external medium density; $E_{k,54}=E_{{\rm k,iso}}/(10^{54}\;{\rm ergs})$ is
the isotropic kinetic energy; $Y$ is the Compton parameter which for fast
cooling
reads $Y \approx [(1+4\epsilon_e/\epsilon_B)^{1/2}-1]/2$,
\citep{Sari01}; $\epsilon_{\rm{e}}$ is the fraction of the internal energy that is
given just behind the shock front to relativistic electrons that form a power-law
distribution of energies: $N_{\rm e}\propto \gamma_{\rm e}^{-p}$ for 
$\gamma_{\rm max}>\gamma_{\rm e}>\gamma_{\rm min}$. 
Assuming equipartition ($\epsilon_e = \epsilon_B=1/3$), $Y \approx 0.62$, Eq.
\ref{eq_cdition_nuc<nuopt} translates into:
\begin{equation} \label{eq_cdition_nuc<nuopt2}
n_0 \gtrsim  2 \times 10^{-3} E_{k,54}^{-1/2}
\end{equation}
For an efficiency of conversion of the kinetic to gamma-rays energy
$\epsilon_{\gamma}=1\%$ the observed $E_{\rm \gamma,iso}=1.1\times
10^{53}\,\rm erg$
(see Sect. \ref{SubSec:specBAT}) implies:
$n_0 \gtrsim  6 \times 10^{-4}$.

Using the best fit simple power-law
models for the break frequency evolution with time of Sect. \ref{subsec:SED}
we have  $\nu_{\rm b}(112\,\rm ks)\sim 1.5\times 10^{14}\,\rm Hz$. Following
\cite{Granot02}, their Table 2, spectral break 9, this means (a value 
that roughly agrees with the results for a range of values for $p$ 
derived below is adopted):
\begin{equation} \label{eq_cdition_num_sim_nuopt_p}
\left(\frac{\bar{\epsilon}_e}{\xi_e}\right)^2 \epsilon_B^{1/2}
\sim 2 \times 10^{-3} \, E_{k,54}^{-1/2}   
\end{equation}
where $\bar{\epsilon}_e = \epsilon_e\gamma_m/\langle\gamma_e\rangle$ and
$\xi_{\rm{e}}$ is the fraction of accelerated electrons.
The value of $p$ is $p=1.97 \pm0.03$ with intrinsic reddening 
$E(B-V)_{z}=0.03$ ($\chi^{2}/\rm{dof}=135/138$). Freezing the intrinsic reddening to 
$E(B-V)_{z}=0.06$ gives $p=2.03\pm 0.02$ ($\chi^{2}/\rm{dof}=140.8/139$) while freezing 
it to $E(B-V)_{z}=0.08$ gives $p=2.08\pm0.02$ ($\chi^{2}/\rm{dof}=158.6/139$). We thus 
take $p=2.0\pm 0.1$. In particular, we calculate the range of values obtained 
for the microphysical parameters in the three cases $p=2.1$, $p=2$ and $p=1.9$ 
since the expression of 
$\bar{\epsilon}_e$ changes when $p>2$, $p=2$ and $p<2$ \citep{Granot06b}:
\begin{eqnarray} \label{eq_bar_epsilon_e}
\frac{\bar{\epsilon}_e}{\epsilon_e} = \left\{ \begin{array}{ll}
\approx (p-2)/(p-1)                                & p>2\\
1/\ln(\gamma_{\rm max}/\gamma_{\rm min})               & p=2\\
(2-p)/(p-1) (\gamma_{\rm min}/\gamma_{\rm max})^{2-p}   & p<2
\end{array} \right. 
\end{eqnarray}
$\gamma_{\rm max}$ is obtained by equating the acceleration and cooling 
times of an electron, and is $\gamma_{\rm max} = \sqrt{3q_e/(\sigma_T B' (1+Y))}$. 
Calculating the magnetic field value by $B' = \gamma_{\rm aft} c \sqrt
{32\pi\epsilon_{\rm B}n m_p}$ and assuming $n_0=1$, $\epsilon_e = 
0.3$, $\epsilon_B = 0.1$ and $\gamma_{\rm aft}=30$ we obtain $\gamma_{\rm max} 
\sim 10^7$. Taking $\gamma_{\rm min} \sim 500$ (obtained for $p\sim 2.1$), 
$(\gamma_{\rm min}/\gamma_{\rm max}) \sim 5 \times 10^{-5}$ (given the way 
this ratio appears in equation (\ref{eq_bar_epsilon_e}) - either in a 
logarithm or with a power $2-p=0.1$ in our case - the dependence of 
the ratio $\bar{\epsilon}_e/\epsilon_e$ on it is very weak, and variations 
in its value have only a small effect). Then, since for $p=2.1$, $(p-2)/(p-1)
\sim 0.1$, and for $p=2$, $1/\ln(\gamma_{\rm max}/\gamma_{\rm min}) \sim 0.1$, for
 $p\ge2$ we obtain $(\epsilon_e/\xi_e)^2 \epsilon_B^{1/2} \sim 0.2$. From the 
equipartition value - giving the maximum possible values  $\epsilon_e/\xi_e= 
\epsilon_B=1/3$ - we obtain an upper limit on the fraction of accelerated 
electrons: $\xi_e \lesssim 0.3$. For $p=1.9$ we have $(2-p)/(p-1) (\gamma_{
\rm min}/\gamma_{\rm max})^{2-p}\sim 0.04$, and then $(\epsilon_e/\xi_e)^2 
\epsilon_B^{1/2} \sim 1.25$, and then $\xi_e \lesssim 0.2$. The constraint 
on the microphysical parameters being very close in all cases, the exact 
value of $p$ is then not of primary importance and the approximation $p=2.0\pm0.1$ 
is then consistent.

The evolution of the peak frequency being consistent with an off-axis 
interpretation of the afterglow, we further test this scenario by 
deriving the viewing and half-opening angle of the jet. The jet break
time is given by \cite{Sari99} for the ISM and \cite{Chevalier00}
for the wind environments:
\begin{equation}
t_{\rm jet} \approx \left\{ \begin{array}{ll}
1.2\; (1+z)\left(\frac{E_{54}}{n_0}\right)^{1/3}
\left(\frac{\Delta \theta}{0.1} \right)^{8/3}\;{\rm days} & (k=0)\ \\
\\
6.25\; (1+z)\left(\frac{E_{54}}{A_*}\right)
\left(\frac{\Delta \theta}{0.1} \right)^{4}\;{\rm days} & (k=2)\
\end{array}\right.
\end{equation}
From Table \ref{Tab:XRTrebfit} we read a post-break power-law decay index
$b=2.1\pm0.1$ ($e=2.3\pm0.1$) if $t_{\rm jet}\sim t_{\rm peak}$
($t_{\rm jet}=t_{\rm br_{2}}$). Both are consistent with being post-jet
break decay indices. We therefore
conservatively assume $t_{\rm jet}<1\,\rm day$, which leads to:
\begin{equation}\label{theta2}
\Delta \theta <
\left\{ \begin{array}{ll}
0.055 \left( \frac{E_{54}}{n_0} \right)^{-1/8}\;{\rm rad} & \quad (k=0)\ \\
\\
0.045 \left( \frac{E_{54}}{A_*} \right)^{-1/4}\;{\rm rad} & \quad (k=2)\
\end{array} \right.
\end{equation}
Evaluating Eq. 9 of \cite{Nousek06} at $t=t_{\rm peak}$, when
$\gamma\sim 1/\theta$ we obtain:
\begin{equation}\label{theta1}
\frac{1}{\gamma(t_{\rm peak})}\approx\theta =\left\{ \begin{array}{ll}
0.03\left(\frac{E_{54}}{n_0} \right)^{-1/8}\rm{rad} &\quad (k = 0)\ \\
\\
0.03 \left(\frac{E_{54}}{A_*}\right)^{-1/4}\rm{rad} &\quad (k = 2)\
\end{array} \right.
\end{equation}
Using Eq. \ref{eq_cdition_nuc<nuopt2} for the ISM environment we
finally have $\theta>0.014E_{k,54}^{-3/16}\,\rm rad$. From the
comparison of Eq. \ref{theta1} and Eq. \ref{theta2} it is apparent that
$\theta>\Delta\theta/2$. Moreover, the slope of the rising part 
of the re-brightening of the afterglow is $\sim 1.8$, which is 
in rough agreement with the rising slope of the re-brightening obtained 
from model 3 of \cite{Granot02b} - see their Fig. 2 - for 
$\theta \sim 3 \Delta \theta$. This is consistent with $\theta> \Delta \theta/2$.

The off-axis interpretation implies that the value
of the observed gamma-ray isotropic energy $E_{\gamma,\rm{iso},\theta}$ corresponds
to an actual on-axis input of $E_{\gamma,\rm{iso},0} \approx a^{-2}E_{\gamma,
\rm{iso},\theta}$ if $\theta<\Delta \theta$
and $E_{\gamma,\rm{iso},0} \approx a^{-3}E_{\gamma,\rm{iso},\theta}$ if $\theta>\Delta \theta$.
Since $E_{\rm \gamma,iso,\theta} \sim 10^{53}\;$erg, 
this may lead to very high energy output for this burst, which may 
be unphysical. It is therefore important to obtain limits on the 
Lorentz factor of the prompt emission, since 
$a^{-1} \approx 1+\Gamma_0^2\theta^2$.
Lower limits to $\Gamma_0$ can be obtained following \cite{Lithwick01}, 
requiring the medium to be optically thin to annihilation of photon pairs
(Eq. \ref{eq_gam_mins_theta_num})
and to scattering of photons by pair-created electrons and positrons
(Eq. \ref{eq_gam_mins_theta_num2})\footnote{See Appendix \ref{App1}
for a complete derivation of Eq. \ref{eq_gam_mins_theta_num} and
\ref{eq_gam_mins_theta_num2}.}:

\begin{eqnarray}\label{eq_gam_mins_theta_num}
\Gamma_{{\rm min, \gamma \gamma}} =
\frac{\widehat{\tau}_{\theta}^{1/(2\beta_{\rm B}+2)}\left(\frac{150\;{\rm
keV}}
{m_e c^2}\right)^{(\beta_{\rm B}-1)/(2\beta_{\rm B}+2)}}{(1+z)^{(1-\beta_{\rm B})/(\beta_{\rm B}+1)}}
\nonumber\\
\times\left\{ 
\begin{array}{ll}
a^{-1/2}   & \theta <\Delta \theta\\
\left(a_*\right)^{1/(2\beta_{\rm B}+2)} a^{-(\beta_{\rm B}+2)/(2(\beta_{\rm B}+1))}
     & \theta > \Delta \theta
\end{array}\right. \
\end{eqnarray}

\begin{eqnarray}\label{eq_gam_mins_theta_num2}
\Gamma_{{\rm min,e^{\pm}}} = \widehat{\tau}_{\theta}^{1/(\beta_{\rm B}+3)}
 (1+z)^{(\beta_{\rm B}-1)/(\beta_{\rm B}+3)}
\,\,\,\,\,\,\,\,\,\,\,\,\,\,\,\,
\nonumber \\
\,\,\,\,\,\,\,\,\,\,\,\,\,\,\,\,\times
\left\{ \begin{array}{ll}
a^{-2/(\beta_{\rm B}+3)}   & \theta < \Delta \theta\\
\left(a_*\right)^{1/(\beta_{\rm B}+3)} a^{-3/(\beta_{\rm B}+3)}
& \theta > \Delta \theta
\end{array}\right. \
\end{eqnarray}
where $\beta_{\rm B}$ is the high energy photon index of the prompt
Band spectrum.

From \cite{Blandford76}, the Lorentz factor at the deceleration 
radius and at the peak of the re-brightening can be related by $\gamma
(R_{\rm peak}) = \gamma(R_{\rm dec})(R_{\rm peak}/R_{\rm dec})^{-(3-k)/2}$. 
The Lorentz factor at the deceleration radius is a factor $g<1$ of 
the Lorentz factor of the prompt emission $\Gamma_0$. Combining this
with $a^{-1} = 1+\Gamma_0^2\theta^2$ and $\theta=1/\gamma(t_{\rm peak})$, 
we obtain the following expression for the parameter $a$:
\begin{equation}
a^{-1} = 1+ g^{-2}\left(\frac{R_{\rm peak}}{R_{\rm dec}}\right)^{3-k}\ .
\end{equation}
Since $g \lesssim 1/2$, and $R_{\rm dec} \lesssim R_{\rm peak}$,
we have $a^{-1} \gtrsim 5$ which, when substituted 
in equation \ref{eq_gam_mins_theta_num} and \ref{eq_gam_mins_theta_num2} and keeping 
the strongest constraint, implies $\Gamma_0 \gtrsim 46$. To consider the 
other extreme case, where the deceleration time is $\sim T_{\rm GRB}$, 
one should be careful in translating the ratio of radii to ratio of times: 
for a prompt emission with a single pulse, the duration of the GRB 
$T_{\rm GRB}$ is the duration of the pulse, which changes with the 
parameter $a$ from on-axis to off-axis, as then does $t_{\rm dec}$. 
We can therefore use off-axis values of the time $t\sim R$ which means
$R_{\rm peak}/R_{\rm dec} \sim t_{\rm peak}/t_{\rm dec} \sim t_{\rm peak}/
T_{\rm GRB}$ in our case here. Since $t_{\rm peak} \sim 2\times10^{4}\;$s 
and $T_{\rm GRB}=264.3\;$s (we identify the duration of the GRB with the
$T_{90}$ parameter), $a^{-1} \gtrsim 300$ for $k=2$ (then 
$\Gamma_0 \gtrsim 230$) and $a^{-1} \gtrsim 1.7\times10^{3}$ ($\Gamma_0 
\gtrsim 17\times10^3$) for $k=0$. In the case of a prompt emission with 
several pulses, as it is the case for GRB081028, each pulse duration 
increases by a factor $a^{-1}$ from on-axis to off-axis, however the 
total duration of the burst does not increase much, approximately by 
a factor of order unity, since the enlargement of pulses is somewhat 
cancelled by their overlapping. In this case, the GRB duration to 
consider is the on-axis one, for which $t\propto R/\gamma^2\propto R^{4-k}$; 
since $t_{\rm peak}$ is the limit between the on-axis and off-axis 
cases we can use $t_{\rm peak}\propto R_{\rm peak}^{4-k}$ and then 
$a^{-1} = 1+ g^{-2}\left(\frac{t_{\rm peak}}{T_{\rm GRB}}\right)^{
(3-k)/(4-k)} \gtrsim 100$ (or $\Gamma_0 \gtrsim 136$) for $k=0$ 
and $a^{-1} \gtrsim 36$ (or $\Gamma_0 \gtrsim 94$) for $k=2$.

The lower limit on the value of $a^{-1}$ thus ranges between\footnote{Since GRB\,081028 is composed of at least two pulses, 
we consider the most relevant case, when the observed off-axis duration 
of the prompt emission is close to the on-axis one.}  $\sim 5$ 
and $\sim 10^2$: this implies values 
of the isotropic on-axis gamma-ray energy output to range between 
$E_{\rm \gamma,iso,0} \sim 3\times10^{54}\;$erg and $E_{\rm \gamma,iso,0} \sim 10^{57}\;$erg 
if $\theta < \Delta \theta$  and even greater values for $\theta > \Delta \theta$: 
between $E_{\rm \gamma,iso,0} \sim 1.4\times10^{55}\;$erg and $E_{\rm \gamma,iso,0} \sim 10^{59}\;$erg. 
These very high values could suggest that the observed prompt emission is from a
different component than the observed afterglow emission. This possibility independently 
arises from the prompt efficiency study: the next section is dedicated to an investigation
of this topic.

\subsection{Prompt efficiency}
\label{subsec_prompt_eff}
The study of the efficiency of the conversion of  the total initial energy
into gamma-rays can in principle shed light on the physical mechanism at 
work. In the particular case of GRB\,081028, this study helps us to understand
if the prompt and afterglow emission originated from physically different
regions. The first part of this sub-section is dedicated to the on-axis case;
the second part to the off-axis case.

Assuming that all energy not radiated in gamma-rays
ends up in the kinetic energy $E_{\rm{k}}$ of the afterglow,
the important parameters are the energy radiated in gamma-rays, 
$E_{\gamma}$, the kinetic energy of the afterglow, $E_{\rm{k}}$ and 
a parameter 
$f\equiv E_{\rm k}(10\;{\rm{hr}})/E_{\rm k,0}$ (\citealt{Granot06b}; 
\citealt{Fan06}, hereafter FP06) that accounts for energy injection 
during the shallow decay phase (since energy injection is the most common 
explanation for this phase), where $E_{\rm k,0}$ is the initial kinetic 
energy of the afterglow, before energy injection. Accounting for
energy injection, the efficiency of the 
prompt emission reads:
\begin{equation}
\epsilon_{\gamma}\equiv \frac{E_{\gamma}}{E_{\rm k,0}+E_{\gamma}} = \frac{f\tilde{\epsilon_{\gamma}}}{1+(f-1)\tilde{\epsilon_{\gamma}}}
\end{equation}
where $\tilde{\epsilon_{\gamma}} \equiv E_{\gamma}/(fE_{\rm k,0}+E_{\gamma})
=E_{\gamma}/(E_{\rm k}(10\;{\rm{hr}})+E_{\gamma})$ is the prompt efficiency in 
the case of no energy injection. All the listed 
quantities are isotropic equivalent quantities.
The value of this parameter can be calculated 
with a good estimate of $E_{\rm k}(10\;{\rm{hr}})$ that can be obtained from 
the X-ray luminosity at 10 hours if the X-ray frequency $\nu_X$ is above 
both $\nu_m$ and $\nu_c$ (FP06; \citealt{Lloyd04}, hereafter LZ04). 
This is the case for GRB\,081028 (see Sect. \ref{subsec_aftmod_offaxis})
which shows an isotropic X-ray luminosity of $L_{\rm{x,iso}}(\,\rm 10\,hr,\rm{obs})\sim(6.3\pm1.0)
\times 10^{47}\;\rm{erg\,s^{-1}}$. The calculation of the kinetic energy is done 
following the prescriptions of FP06: unlike LZ04, they integrate their model over 
the observed energy band $0.3-10\;$keV and consider the effect of inverse Compton 
cooling. Equation (9) of FP06 gives the kinetic energy at ten hours:
\begin{eqnarray}
\label{Eq:Ek}
E_{\rm k}(10\;{\rm{hr}})=R\,L_{X,46}^{4/(p+2)}\Big(\frac{1+z}{2}\Big)^{(2-p)/(p+2)}\times
\nonumber
\\\epsilon_{\rm{B,-2}}^{-(p-2)/(p+2)}
\epsilon_{\rm{e,-1}}^{4(1-p)/(p+2)}
(1+Y)^{4/(p+2)}
\end{eqnarray}
where $R=9.2\times 10^{52}[t(10\,\rm{hr})/T_{90}]^{\frac{17\epsilon_{e}}{16}}\,\rm{erg}$.
This implies we need to make some assumptions on the microphysical parameters 
$\epsilon_e$, $\epsilon_B$ and $Y$. For the 
latter, as the afterglow is likely to be in fast cooling 
(see Sect. \ref{subsec_aftmod_offaxis}), then $Y>1$ and we take $Y \sim 
(\epsilon_e/\epsilon_B)^{1/2}$ following FP06. \cite{Medvedev06} showed that during the 
prompt emission it is most likely that $\epsilon_e\approx\sqrt{\epsilon_B}$. The 
values of the microphysical parameters being poorly constrained 
(Sect. \ref{subsec_aftmod_offaxis}), we set $\epsilon_e=0.3$ and $\epsilon_B=0.1$, 
which is consistent with the values obtained in subsection \ref{subsec_aftmod_offaxis}
(see eq. \ref{eq_cdition_num_sim_nuopt_p} when $\xi_e <1$ and eq.
\ref{eq_bar_epsilon_e} and the paragraph below it). 
Taking $p\sim2$, we thus obtain 
$E_{\rm k}(10\;{\rm{hr}}) = 1.3\times10^{55}\;$erg. Combined with the 
observed isotropic gamma-ray energy of the prompt emission 
$E_{\rm \gamma} = 1.1\times10^{53}\;$ergs, 
we have
$\tilde{\epsilon_{\gamma}} = 8.6\times10^{-3}$ (corresponding to a ratio $E_{\rm k}(10\;{\rm{hr}})/E_{\rm \gamma} \approx 116$): 
this is low, even compared to the values obtained by FP06 (their values being between 
$0.89$ and $0.01$ - see their Table 1), which are already lower than previous estimates 
by LZ04. Now returning to the efficiency including energy injection, we can obtain an
estimate of  $E_{\rm k,0}$ by using the previous formula but at the peak of 
the re-brightening and taking $R=9.2\times 10^{52}$  erg (thus ignoring 
energy radiative losses since the end of the prompt emission), which with its peak 
luminosity $L_{\rm{peak}} = 1.2 \times 10^{48}\;$erg s$^{-1}$ gives an initial kinetic 
energy injected into the afterglow  $E_{\rm k,0} = 1.16 \times 10^{55}\;$erg and 
then an efficiency of the prompt emission which is as low as
 $\epsilon_{\gamma}= 9.4\times 10^{-3}$.
This calculation assumes an on-axis geometry and accounts for energy injection.

\begin{figure}
\centering
    \includegraphics[scale=0.44]{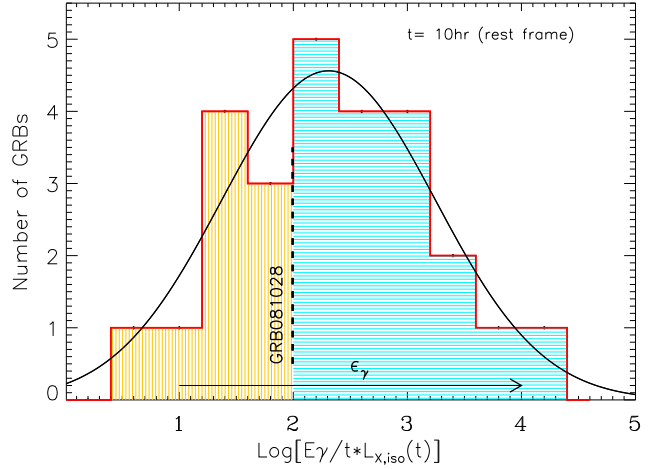}
      \caption{Distribution of $E_{\gamma}/t L_{X}(t)$ with $t=10\,\rm{hr}$
rest frame, for the sample of 31 long GRBs detected by \emph{Swift} with $E_{\gamma}$ provided by Amati et al. (2008). 
Black solid line: Gaussian best fit to the distribution. The dashed black line marks the position of GRB\,081028 in the distribution, while the
black solid arrow is pointed to the direction of increase of the radiative efficiency
parameter $\epsilon_{\gamma}$.}
\label{Fig:EisoLisoRest}
\end{figure}

To strengthen the result of the above paragraph that the efficiency of 
GRB\,081028  when considered on-axis is low, we analyse its prompt 
and afterglow fluencies and compare them to a sample of Swift bursts from 
\cite{Zhang07b}, since fluences require no assumptions to be obtained.
The prompt $1-10^4\,\rm{keV}$ 
gamma-ray fluence\footnote{Depending on the high energy
slope of the Band spectrum, we have  $S_{\rm \gamma}\sim 6.6\times 10^{-6}\,\rm{erg\,
cm^{-2}}$ for $\beta_{\rm B}= -2.5$ and $S_{\rm \gamma}\sim 9.5\times 10^{-6}\,\rm{erg\,
cm^{-2}}$ for $\beta_{\rm B}= -2.1$.} of GRB\,081028 is $S_{\rm \gamma}\sim 8\times 10^{-6}\,\rm{erg\,
cm^{-2}}$ and its afterglow X-ray fluence, calculated by $S_{\rm X} \sim t_{\rm peak} F_{\nu}
(t_{\rm peak})$ to be consistent with \cite{Zhang07b} 
method, is $S_{\rm X} \approx 3\times 10^{-7}\rm{erg\,cm^{-2}}$, so that their 
ratio is $S_{\gamma}/S_{\rm{X}} \approx 26.7$, placing GRB\,081028 in the lower 
part of Fig. 6 of \cite{Zhang07b}. Compared to their sample of 31 {\it Swift} bursts, the $15-150\;$
keV fluence of GRB081028, which is $3.2\times 10^{-6}\rm{erg\,
cm^{-2}}$ is well within their range of values (spanning from $S_{\rm X,min} \approx 
8\times 10^{-8}\rm{erg\,cm}^{-2}$ to $S_{\rm X,max} \approx 1.5\times 10^{-5}\rm{erg\,cm}^{-2}$; sixth column 
of their table 1), whereas its X-ray fluence is higher than most of them (see columns 6-9 of their 
table 2). It thus means that whereas GRB\,081028 released as much energy in its prompt 
emission as most bursts, more kinetic energy was injected in its outflow. 
This gives a lower efficiency than most of the GRBs analysed by \cite{Zhang07b},
consistent with the scenario above.
Figure \ref{Fig:EisoLisoRest} clearly shows that this is likely 
to be extended to other \emph{Swift} long GRBs: at late afterglow epoch the X-ray band 
is above the cooling frequency and the X-ray luminosity is a good probe of 
the kinetic energy.
In particular $E_{\rm{k}}\propto L_{\rm{x,iso}}$ (see Eq. \ref{Eq:Ek}): this means that high (low)  
values of the ratio $E_{\gamma}/L_{\rm{x,iso}}$ are linked to high (low) values of 
radiative efficiency.

The afterglow modelling of the previous section favours an off-axis
geometry. In this case, considering that $E_{\rm k,iso} \sim 10^{55}\;$erg, for the lower 
limit $a^{-1}\sim 5$ (see Sect. \ref{subsec_aftmod_offaxis}) the efficiency of the prompt emission 
becomes $\epsilon_{\gamma} \sim 0.23$, which is a more usual value 
(it is in the middle of the efficiency distribution of FP06). 
However, the upper limit of the range of values for $a^{-1}$ gives an 
efficiency of $99\%$  (when $\theta < \Delta \theta$, and thus an even 
higher value for $\theta > \Delta \theta$), which is exceptionally high and 
very hard to 
reconcile with models of the prompt emission. This would suggest that 
the observed prompt emission is from a different component than the 
observed afterglow emission.

An alternative way of achieving a more reasonable gamma-ray efficiency is if 
the observed prompt gamma-ray emission is from material along our line of sight, 
which has $E_{\rm{k,iso}} \sim E_{\rm{\gamma,iso}}$, while the peak in the X-ray and optical 
light-curves at $\sim2\times 10^4\;$s is from a narrow jet-component pointed away from 
us that has a significantly higher $E_{\rm{k,iso}}$. In this picture the afterglow 
emission of this material along our line of sight (and possibly also between 
our line of sight and the core of the off-axis jet component) could account for 
the very flat (almost constant flux) early optical emission (from the white 
light detection at $275\;$s, through the $R$-band detection at $1780\;$s, and 
the I-band detections at several thousand seconds). This early optical emission 
appears to be from a different origin than the contemporaneous X-ray emission, 
and is most likely afterglow emission, regardless of the origin of the prompt emission:
the observed X-ray and optical emission in the time interval 
$1.8\,\rm{ks}\leq t \leq9.5\,\rm{ks} $ implies a spectral index $|\beta_{OX}|<0.5$.
Conversely, assuming $\beta_{OX}=0.5$, the expected X-ray contribution of the on-axis 
component at these times is $\approx 3\times 10^{-4}\,\rm{mJy}$ which is lower than the 
observed X-ray flux for $t<9\,\rm{ks}$ and comparable to the observed one at 
$t\sim 9\,\rm{ks}$.

\section{Summary and conclusions}
\label{sec:conclusion}

The 0.3-10 keV X-ray emission of GRB\,081028 consists of 
a flat phase up to 
$\sim300$ s (the XRT is likely to have captured the prompt emission in the
X-ray energy band) followed by a steep decay with flares superimposed extending 
to $\sim7000$ s (component 1). The light-curve then shows a re-brightening which
starts to rise at $t\sim 8000\,\rm s$ and
peaks around $20\,\rm ks$ (component 2). The different spectral and temporal 
properties strongly characterise the XRT signal as due to two distinct 
\emph{emission} components.
However, their further characterisation as emission
coming from \emph{physically} distinct \emph{regions} is model dependent.

The strong hard-to-soft evolution characterising the prompt and steep
decay phase of GRB\,081028 from trigger time to 1000 s is 
well modelled by a shifting Band function: the spectral peak energy
evolves to lower values, decaying as $E_{\rm peak}\propto t^{-7.1\pm0.7}$ 
or $E_{\rm peak}\propto (t-t_0)^{-4.2\pm2.4}$ when the zero-time of the 
power-law is allowed to vary: the best fit constrains 
this parameter to be $t_0=109\pm89$ s. In either case our results
are not consistent with the $\propto t^{-1}$ behaviour
predicted by the HLE in its simplest formulation. While a more
realistic version of this model might still account for the 
observed $E_{\rm peak}$ evolution, other possibilities must be 
investigated as well: 
the adiabatic expansion cooling of the $\gamma$-ray source predicts
a steeper than observed light-curve decay and is therefore unlikely.
While the peak is moving, a softening of both the low and high-energy
portions of the spectrum is clearly detected. 
The failure of both the curvature effect and the adiabatic cooling argues 
against the abrupt switch-off of the GRB source after the prompt emission 
and suggest the continuation of the central engine activity during the
steep decay. An off-axis explanation 
may reconcile the high latitude emission or the adiabatic expansion 
cooling models with the data. This will be explored in a future work.

GRB\,081028 has afforded us the unprecedented opportunity to track a
smoothly rising X-ray afterglow after the steep decay: the rising
phase of the emission component later accounting for the shallow 
light-curve phase is usually missed, being hidden by the steep 
decay which is the tail of the prompt emission both from the
spectral and from the temporal point of view. The peculiarity
of GRB\,081028 lies in a small overlap in time between the steep
decay and the following re-brightening caused by an unusual delay 
of the onset of the second component of emission. Contemporaneous optical 
data allow the evolution of the SED during the re-brightening to be constrained:
the spectral distribution is found to be best described by a
photo-electrically absorbed smoothly broken power-law with a break
frequency evolving from $1.6\times 10^{15}\,\rm Hz$ downward to the
optical band. The break frequency can be identified with the 
injection frequency of a synchrotron spectrum in the fast cooling regime
evolving as $\nu_{\rm b}\propto t^{-2.6\pm0.2}$. 
The intrinsic optical absorption is found to satisfy $A_{V,z}<0.22$.

The observed break frequency scaling is inconsistent with the 
standard predictions of the onset of the forward shock emission even if 
this model is able to account for the temporal properties
of the X-ray re-brightening (note that in this context 
the delay of the second emission component is due to a lower
than usual fireball Lorentz factor or external medium density).
Alternative scenarios have therefore 
been considered. While a dust scattering origin 
of the X-ray emission is ruled out since we lack 
observational evidence for a non-negligible dust extinction and
strong spectral softening, a reverse shock origin cannot be 
excluded. However, this can be accomplished only by requiring
non-standard burst parameters: the ejecta should have a tail of 
Lorentz factors decreasing to low values; $\epsilon_{e}$ should be
near equipartition; only a small fraction $\xi_{e}\sim 10^{-2}$
of electrons should contribute to the emission. 

The predictions of the off-axis model have been discussed in detail:
according to this model a peak of emission is expected when the 
beaming cone widens enough to engulf the line of sight. 
The delayed onset of the second emission component is not a consequence 
of unusual intrinsic properties of the GRB outflow but is 
instead an observational artifact, due to the off-axis condition.
The observed evolution of $\nu_{\rm b}$ is consistent with the
expected evolution of the injection frequency of a fast cooling 
synchrotron spectrum for $0\lesssim k\lesssim2$. We interpret the light-curve
properties as arising from an off-axis view, with $\theta\sim3\Delta
\theta$ and $\theta\sim0.03(\frac{E_{54}}{n_{0}})^{-1/8}$ for $k$=0 
(or $\theta\sim0.03(\frac{E_{54}}{A_{*}})^{-1/4}$ for $k$=2), 
$\theta$ being the angle from the outer edge of the jet and $\Delta\theta$
the jet opening angle. 

In this scenario, the peculiarity of GRB\,081028, or the reason why we do 
not observe more GRB\,081028-like events, may be attributed to the following 
reasons. Since GRB\,081028 is a particularly bright (and therefore rare) 
event when viewed on-axis (with high on-axis $E_{\rm{iso}}$ and $L_{\rm{iso}}$ values), 
it is detectable by an off-axis observer even at the cosmological distance 
implied by its redshift z = 3.038. In addition, GRB\,081028 appears to be 
characterized by a particularly narrow jet, for which the ratio of the 
detectable off-axis solid angle to on-axis solid angle is larger than for 
wider (but otherwise similar) jets. Finally, GRB\,081028 might have a 
peculiar angular structure that is not representative of most GRBs, which 
would undermine the drawing of statistical conclusions under the 
assumption of a similar angular structure for most or all GRB jets.


The radiative efficiency is one of the key parameters in GRB science:
a precise estimate of this parameter would allow one to distinguish 
between different models put forward to explain the observed emission. 
For the on-axis model, with $\epsilon_{\gamma}\sim 10^{-2}$, the GRB\,081028 
efficiency turns out 
to be lower than the values obtained by FP06 and LZ04
for a sample of pre-\emph{Swift} GRBs:
this directly implies that instead of having released as much energy
in the prompt emission as most bursts of the two samples, GRB\,081028
has a much greater kinetic energy injected in the outflow.
Figure \ref{Fig:EisoLisoRest} clearly shows that this conclusion is 
likely to be extended to other \emph{Swift} bursts with secure 
$E_{\gamma,\rm{iso}}$ measurement. This picture changes if we 
consider the off-axis interpretation: if the deceleration time is much 
longer than the prompt duration  the 
prompt and afterglow emission are consistent with originating from 
the same physical component and the efficiency of the burst is 
comparable to most bursts; if instead the deceleration time is close to 
the end of the prompt emission, then the on-axis isotropic energy output 
would imply an extremely high efficiency of $99\%$ which is very hard to 
explain. This suggests that the prompt and afterglow emission come 
from different physical components.


GRB\,081028 demonstrates the evolution of GRB spectral properties
from the onset of the explosion to $\sim10^{6}\,\rm s$ after trigger 
and shows that this is likely to be attributed to two distinctly 
contributing components of emission. These can be constrained only by prompt,
broad-band coverage and good time resolution observations.

\section*{Acknowledgments}
J.G. gratefully acknowledges a Royal Society Wolfson Research Merit Award.
Partly based on observations made with the Nordic Optical Telescope,  
operated on the island of La Palma jointly by Denmark, Finland,  
Iceland, Norway, and Sweden, in the Spanish Observatorio del Roque de  
los Muchachos of the Instituto de Astrofisica de Canarias.
The Dark Cosmology Centre is funded by the Danish National Research  
Foundation. The PAIRITEL work of J.S.B., A.A.M, and D.S. was partially 
supported by NASA grant NNX09AQ66G.
This work is supported by ASI grant SWIFT I/011/07/0, by  the 
Ministry of University and Research of Italy (PRIN MIUR 2007TNYZXL), by
MAE and by the University of Milano Bicocca (Italy).


\newpage
\appendix
\section{Derivation of equations 18 and 19}
\label{App1}
As for the main article, the convention of a subscript 0 ($\theta$) 
for on-axis (off-axis) quantities is used. A subscript $``*"$ is added
for the cases when $\theta = \Delta \theta$, $\Delta \theta$ being the
jet opening angle.
Following \cite{Lithwick01}, their eq. 5 and 8, the lower limit 
to $\gamma_0$ due to photons annihilation reads:
\begin{equation}\label{eq_gam_min_gg}
\gamma_{{\rm min, \gamma \gamma}} \equiv \frac{\widehat{\tau_0}^{1/(2\beta_{\rm B}+2)}}
{(1+z)^{(1-\beta_{\rm B})/(\beta_{\rm B}+1)}} \left(\frac{E_{\rm max}}
{m_e c^2}\right)^{(\beta_{\rm B}-1)/(2\beta_{\rm B}+2)} 
\end{equation}
while considering the scattering of photons by pair-created electrons 
and positrons:
\begin{equation}\label{eq_gam_min_pp}
\gamma_{{\rm min,e^{\pm}}} \equiv \widehat{\tau_0}^{1/(\beta_{\rm B}+3)} (1+z)^{(\beta_{\rm B}-1)/(\beta_{\rm B}+3)}\ 
\end{equation}
where: $E_{\rm max}=150$ keV for BAT observations; $\beta_B$ is the high 
energy photon index of the prompt spectrum; $z$ is the redshift of the burst.
From eq. 4 of \cite{Lithwick01}, the dimensionless quantity $\widehat{\tau}$ 
can be re-written as:
\begin{equation}\label{eq_widehattau_f1}
\widehat{\tau_0} = (2.1\times10^{11})\; \frac{(d_L/7\, \rm{Gpc})^2(0.511)^{(1-\beta_B)}f_{1,0}}
{(\delta T_0/0.1\,\rm{s})(\beta_B-1)}\ 
\end{equation}
where: $d_L$ is the luminosity distance; $\delta T_0$ is the typical 
time scale of variability and $f_{1,0}$ is the on-axis number of
photons per second per square centimeter per MeV at the energy of 1 MeV.
The on-axis quantities must be now related to the observed off-axis
ones. In particular from eq. \ref{Eq:offaxis1} directly follows 
$\delta T_0=a\, \delta T_{\theta}$ and $\nu_0 = \nu_{\theta}/a$.
The fluence $\mathcal{F} = \int \frac{dtdE E dN}{dEdAdt} \propto E dN/dA$
while $f=\int \frac{dtdE  dN}{dEdAdt} \propto  dN/dA$.
For a point source located at $\theta>\Delta \theta$ 
$dN/dA\propto d\Omega\propto \delta^2$: this implies 
$\mathcal{F}_{0} = a^{-3} \mathcal{F}_{\theta}$,
$f_{0} = a^{-2} f_{\theta}$. When $\theta<\Delta \theta$ the size of the 
region significantly contributing to the observed emission increases 
as $\theta \propto a^{-1}$: this translates into 
$\mathcal{F}_{0} = a^{-2} \mathcal{F}_{\theta}$, $f_{0} = a^{-1} f_{\theta}$.
From the fact that $\widehat{\tau}\propto \frac{f}{\delta T}$ and
requiring the continuity of the function at $\Delta \theta$, it
follows:
\begin{eqnarray} \label{eq_tau_dep_a}
\widehat{\tau}_0 = \left\{
\begin{array}{ll}
a^{-2} \widehat{\tau}_{\theta} & \theta < \Delta \theta\\
a_*^{-2} \left(\frac{a}{a_*}\right)^{-3} \widehat{\tau}_{\theta} & \theta > \Delta \theta
\end{array} \right. \ 
\end{eqnarray}
where we remind the reader that $a_*\equiv a(\Delta \theta) 
= 1/(1+\Gamma_0^2 \Delta \theta^2)$.\\
Substituting this result into eq. \ref{eq_gam_min_gg}, leads to:
\begin{eqnarray}\label{eq_gam_mins_theta_numappendix}
\Gamma_{{\rm min, \gamma \gamma}} =
\frac{\widehat{\tau}_{\theta}^{1/(2\beta_{\rm B}+2)}\left(\frac{150\;{\rm
keV}}
{m_e c^2}\right)^{(\beta_{\rm B}-1)/(2\beta_{\rm B}+2)}}{(1+z)^{(1-\beta_{\rm B})/(\beta_{\rm B}+1)}} 
\nonumber\\
\times\left\{
\begin{array}{ll}
a^{-1/2}   & \theta < \Delta \theta\\
\left(a_*\right)^{1/(2\beta_{\rm B}+2)}a^{-(\beta_{\rm B}+2)/(2\beta_{\rm B}+2)}  & \theta > \Delta \theta
\end{array}\right .\
\end{eqnarray}

\begin{eqnarray}\label{eq_gam_mins_theta_num2appendix}
\Gamma_{{\rm min,e^{\pm}}} = \widehat{\tau}_{\theta}^{1/(\beta_{\rm B}+3)}
 (1+z)^{(\beta_{\rm B}-1)/(\beta_{\rm B}+3)}
\,\,\,\,\,\,\,\,\,\,\,\,\,\,\,\,
\nonumber \\
\times
\left\{ \begin{array}{ll}
a^{-2/(\beta_{\rm B}+3)}   & \theta < \Delta \theta\\
\left(a_*\right)^{1/(\beta_{\rm B}+3)} a^{-3/(\beta_{\rm B}+3)}
& \theta > \Delta \theta
\end{array}\right. \
\end{eqnarray}
The prompt spectrum of GRB\,081028 does not allow to constrain the high 
energy photon index $\beta_{\rm B}$, being consistent with a cut-off 
power-law (see Table \ref{Tab:Epeak}). Using $\beta_{\rm B}=-2.5$
(value we observe around $600$ s, observer frame),
$f_1=1.6\times10^{-3}\,\rm{photons}\,\rm{cm}^{-2}s^{-1}\,MeV^{-1}$. The observed 
evolution of $\beta_{\rm B}$ (see Sect. \ref{SubSec:Epeak}) implies a harder high energy
spectrum at $t<600$ s: using $\beta_{\rm B}=-2.1$ we have
$f_1=3.6\times10^{-3}\,\rm{photons}\,\rm{cm^{-2}s^{-1}\,\rm{MeV}^{-1}}$. In the 
following $f_1\approx2\times10^{-3}\,\rm{photons}\,\rm{cm^{-2}s^{-1}\,\rm{MeV}^{-1}}$
will be used. 

Equation \ref{theta2} defines an upper limit to $\Delta\theta$ that
translates into a lower limit to $a_*$ considering that 
$a(\theta)\approx/(1+\gamma^2 \theta^2)\approx (\gamma\theta)^{-2}$
for $\gamma\theta\gg 1$. Inserting this information in the
equation above and using $\delta T_{\theta}=70\,\rm s$ (variability 
time associated to the two pulses, Sect. \ref{SubSec:taBAT}), $d_{\rm L}=17.4\, \rm Gpc$, 
$\beta_{\rm{B}}=2.5$,  $\widehat{\tau_\theta}\approx 6.8\times 10^6$,
we finally obtain eq. \ref{eq_gam_mins_theta_num} and eq.
\ref{eq_gam_mins_theta_num2}. 

\section{Tables}
\vskip 5.0 true cm
\setcounter{table}{0}

\begin{table*}\footnotesize
\begin{center}
\begin{tabular}{lllllllll}
\hline
\bf Tmid & \bf Exp & \bf mag & \bf mag & & & \bf Flux & \bf Flux &  \\
 & &   \bf obs & \bf corr &&& \bf obs& \bf corr&\\
 \bf (s)& \bf (s)& &&&& \bf (mJy)& \bf (mJy)&\\
\hline
WHITE& &    &  &&& & &\\
\hline
275.2	&	147.4	&	20.86    &	20.70&	+0.46&	-0.32&	$8.757\cdot 10^{-3}$&	$1.023\cdot 10^{-2}$	&$\pm3.030\cdot 10^{-3}$\\
663.1	&	19.4	&	$>21.24$ &$>20.87$       & --   & --  &  $<7.480\cdot 10^{-3}$ & $<8.747\cdot 10^{-3}$ &--\\
5174.2	&   196.6	&	$>21.02$ &$>20.85$       &  --  &  -- &	$<7.619\cdot 10^{-3}$&  $<8.910\cdot 10^{-3}$&  --   \\
6580.5	&   139.6	&	$>20.38$ &$>20.21$	     &  --  & --  &  $<1.374\cdot 10^{-2}$ & $<1.606\cdot 10^{-2}$ &  --   \\
101479.4&	8890.0	&    22.94	 &22.77	 &  +1.51&	-0.61&	$1.301\cdot 10^{-3}$&  $1.520\cdot 10^{-3}$	&$\pm9.769\cdot 10^{-4}$\\
124146.5&	8872.8	&    21.46	 &21.29	 &  +0.25&	-0.21&	$5.078\cdot 10^{-3}$&  $5.934\cdot 10^{-3}$	&$\pm1.056\cdot 10^{-3}$\\
\hline
V& &    &  &&& & &\\
\hline
185.9   &	9.1	&	    $>16.99$ &$>16.88$       &  --  & --  &  $<5.811\cdot10^{-1}$&   $<6.430\cdot10^{-1}$&  --\\      	
366.6	&	19.5&		$>18.60$ &$>18.49$	     &  --  & --  &  $<1.319\cdot10^{-1}$&   $<1.460\cdot10^{-1}$& --\\            	
712.7	&	19.4&		$>18.06$ &$>17.95$	     &  --  & --  &  $<2.169\cdot10^{-1}$&   $<2.400\cdot10^{-1}$ & --\\      
4149.8	&   196.6&		$>18.86$ &$>18.75$	     &  --  & --  &  $<1.038\cdot10^{-1}$&   $<1.149\cdot10^{-1}$ &  --  \\  
5584.7	&   196.6&		$>19.63$ &$>19.52$	     &  --  &  -- &  $<5.108\cdot10^{-2}$&   $<5.6524\cdot10^{-2}$&  --  \\  
11192.6	&   598.5&		20.64    &20.54 &+0.53 &-0.35&    $2.016\cdot10^{-2}$&    $2.225\cdot10^{-2}$&   	     $\pm7.752\cdot10^{-3}$  \\ 
28542.1	&   598.5&		19.82	 &19.71 &+0.23 &-0.19&    $4.304\cdot10^{-2}$&    $4.750\cdot10^{-2}$&      	$\pm8.264\cdot10^{-3}$    \\
45891.3	&   598.6&		19.39	 &19.28 &+0.16 &-0.14&    $6.392\cdot10^{-2}$&   $7.054\cdot10^{-2}$&     	$\pm8.643\cdot10^{-3} $   \\
57502.6	&   598.6&		19.51	 &19.40 &+0.22 &-0.18&    $5.745\cdot10^{-2}$&    $6.341\cdot10^{-2}$&    	$\pm1.043\cdot10^{-2}$    \\
101806.7&	8961.6&	    $>19.94$	 &$>19.83$	         &--& --  &   $<3.839\cdot10^{-2}$&   $<4.249\cdot10^{-2}$&  --  \\    
156450.1&	6146.6&	    $>19.65$	 &$>19.55$	         &  --  & --  &   $<5.015\cdot10^{-2}$&   $<5.498\cdot10^{-2}$& --   \\    
\hline
B& &    &  &&& & &\\
\hline
465.7    &19.4&		    $>18.77$ &$>18.63$      &  --  & --  &	  $<1.261\cdot10^{-1}$& $<1.435\cdot10^{-1}$&  -- \\     
4969.6	 &196.6&		$>21.54$ &$>21.40$      &  --  & --  &	  $<9.837\cdot10^{-3}$& $<1.119\cdot10^{-2}$&  --   \\ 
6404.7	 &196.6&		$>20.66$ &$>20.53$      &  --  & --  &	  $<2.212\cdot10^{-2}$& $<2.493\cdot10^{-2}$&   --\\   
17796.8	 &506.1&		20.34	 &20.20 &+0.23 &-0.19&	 $2.971\cdot10^{-2}$&    	$3.371\cdot10^{-2}$&    	$\pm5.730\cdot10^{-3}$   \\ 
35208.9	 &483.4&		19.92	 &19.78 &+0.18 &-0.15&	 $4.377\cdot10^{-2} $ &  	$4.966\cdot10^{-2}$&    	$\pm6.686\cdot10^{-3}$   \\ 
64072.7	 &474.4&		20.80	 &20.66 &+0.38 &-0.28&	 $1.943\cdot10^{-2}$&    	$2.205\cdot10^{-2}$&    	$\pm5.705\cdot10^{-3}$   \\ 
101152.1 &8814.4&	    $>21.54$ &$>21.41$      &   -- & --  &$<9.837\cdot10^{-3}$&  $<1.10886\cdot10^{-2}$& --       \\               
155772.9 &6047.3&	    $>22.70$ &$>22.56$      & --   & --  &$<3.380\cdot10^{-3}$&  $<3.845\cdot10^{-3}$&--\\    
\hline
U& &    &  &&& & &\\
\hline
613.8&		19.5&		$>19.24$ &$>19.07$       &  --  & --  &	  $<2.899\cdot10^{-2}$ &           $<3.390\cdot10^{-2}$& --  \\     
16976.9&	598.6&	    $>20.58$ &$>20.42$       &  --  & --  &	  $<8.438\cdot10^{-3}$ &           $<9.778\cdot10^{-3}$& --  \\  
23578.2&	511.4&		$>19.97$ &$>19.80$	     &  --  & --  &	   $<1.480\cdot10^{-2}$&            $<1.731\cdot10^{-2}$& --  \\     
37243.3&	3516.9&	    21.00  &20.83  &+0.38 &-0.28      & $5.764\cdot10^{-3}$ &          	$6.727\cdot10^{-3}$  & $\pm1.693\cdot10^{-3}$  \\         
66660.8&	3671.5&	    $>20.63$ &$>20.47$	     &  --  &--   &	   $<8.058\cdot10^{-3}$ &          $<9.337\cdot10^{-3}$ &  --  \\
123770.1&	8692.4&	    $>21.08 $&$>20.91$	     & --   & --  &	   $<5.324\cdot10^{-3}$&          $ <6.226\cdot10^{-3} $&  -- \\ 
155516.4&	5699.7&	    $>20.48$ &$>20.31 $      &  --  &  -- &	  $  <9.252\cdot10^{-3}$&  	      $<1.082\cdot10^{-2} $& -- \\   
\hline
UVW1& &    &  &&& & &\\
\hline
416.4&		19.5&		$>22.03$&	$>21.80$       &  --  & --  &	  			$<1.412\cdot10^{-3}$    &       $<1.746\cdot10^{-3}$ &-- \\             
589.5&		19.4&		$>18.68$&	$>18.45$       & --   & --  &	   			$<3.090\cdot10^{-2}$     &       $<3.819\cdot10^{-2}$   &--\\             
5994.8&	    196.6&		$>22.63$&	$>22.40$        &  --  & --  &	  			$<8.128\cdot10^{-4}$   &       $ <1.005\cdot10^{-3}$ &  --  \\         
33415.1&	885.6&		$>21.83$&	$>21.60 $       &  --  & --  &	  			$<1.698\cdot10^{-3}$    &       $ <2.099\cdot10^{-3} $ & --   \\         
40105.9&	885.6&		$>23.16$&	$>22.93 $       &  --  & --  &	  			$<4.988\cdot10^{-4} $  &       $ <6.165\cdot10^{-4} $&  --  \\         
51671.4&	885.6&		$>22.86$&	$>22.63 $       &  --  & --  &	  			$<6.576\cdot10^{-4}$   &       $ <8.128\cdot10^{-3} $&  --  \\         
65720.4&	4231.4&	    $>22.19$&	$>21.96$       & --   & --  &	  			$<1.219\cdot10^{-3} $   &       $ <1.506\cdot10^{-3} $  & --     \\      
386961.8&	36503.7&	$>24.37$&	$>24.14$        &  --  & --  &	  			$<1.637\cdot10^{-4}$    &      $ <2.023\cdot10^{-4} $ & --  \\         
733574.4&	42008.6&	$>23.92$&	$>23.69$       &  --  & --  &	  			$<2.477\cdot10^{-4} $   &      $ <3.062\cdot10^{-4} $ & --  \\         
 \hline
UVM2& &    &  &&& & &\\
\hline
564.5&		19.4&		$>19.44$&	$>19.17$       & --   & --  &				$<1.477\cdot10^{-2}$ &           	$<2.663\cdot10^{-3} $  &  --\\      
5789.7&	    196.6&		$>21.30$&	$>21.04$       &  --  &  -- &				$<2.663\cdot10^{-3}$ &          		$<3.383\cdot10^{-3} $ &  -- \\        
29392.5&	771.3&		$>21.80$&	$>21.53$       &  --  & --  &				$<1.680\cdot10^{-3}$ &          		$<2.154\cdot10^{-3} $ &   -- \\        	
54444.4&	4565.2&  	$>23.48$&	$>23.22$      &  --  & --  &				$<3.576\cdot10^{-4}$ &               $  <4.543\cdot10^{-4} $  &   --    \\   	
68159.5&	885.6&		$>21.30$&	$>21.03$       &  --  & --  &				$<2.663\cdot10^{-3} $ &         		$<3.415\cdot10^{-3}  $&   -- \\        
\hline 
\end{tabular}
\caption{continue....}
\end{center}
\end{table*}
\setcounter{table}{0}
\begin{table*}\footnotesize
\begin{center}
\begin{tabular}{lllllllll}
\hline
\bf Tmid & \bf Exp & \bf mag & \bf mag & & & \bf Flux & \bf Flux &  \\
 & &   \bf obs & \bf corr &&& \bf obs& \bf corr&\\
 \bf (s)& \bf (s)& &&&& \bf (mJy)& \bf (mJy)&\\
\hline
UVW2& &    &  &&& & &\\
\hline
515.3&	    19.5&		$>21.77$&	$>21.47 $	     &  --  & --  &		$<1.454\cdot10^{-3}$ &          	$<1.917\cdot10^{-3}$& --\\             
5380.0&	    196.6&		$>21.48$&	$>21.18$      &  --  & --  &			$<1.900\cdot10^{-3} $   &       	$<2.504\cdot10^{-3}$& -- \\            
56593.0&	885.6&		$>20.96$&	$>20.65$ 	     &  --  & --  &		$<3.067\cdot10^{-3} $   &       	$<4.080\cdot10^{-3}$ & --\\            
\hline  
\end{tabular}
\caption{\emph{Swift}-UVOT photometric set of GRB\,081028. $3\sigma$ 
upper limits are provided 
in cases of non-detection. Column 1: observations mid-time since BAT 
trigger; column 2: exposure time; columns 3 and 5: observed magnitudes and fluxes; 
columns 4 and 8: extinction corrected magnitudes and fluxes; columns 6 and 7
report the errors on the extinction corrected magnitudes, while column 9
lists the errors on the extinction corrected flux.
Only the Galactic extinction correction has been applied to the data.}
\label{Tab:UVOTdata}
\end{center}
\end{table*}

\begin{table*}\footnotesize
\begin{center}
\begin{tabular}{lllllll}
\hline
\bf Tmid& \bf Filter &     \bf Exp& \bf  mag    &  \bf Flux   & \bf mag &  \bf Flux\\
 &        &       &\bf obs   & \bf  obs  & \bf corr &\bf corr   \\  
\bf (s) &        &     \bf (s) &  & \bf (mJy)  &  & \bf (mJy)  \\
\hline  
CrAO &  &    &      &     & &  \\
\hline
  1779.84&	   R&       23x60&     $21.62  \pm   0.07$&    $(6.922  \pm  0.446)\cdot10^{-3}$ & $21.545 \pm   0.07$  &     $ (7.418 \pm  0.478)\cdot10^{-3}$\\
  3585.60&     I&       30x60&     $21.32   \pm  0.09$&    $(7.560  \pm  0.627)\cdot10^{-3}$ & $21.264 \pm   0.09$   &    $ (7.961  \pm 0.660)\cdot10^{-3}$\\
  5529.60&	   I&       30x60&     $21.43   \pm  0.09$&    $(6.832  \pm  0.566)\cdot10^{-3}$ & $21.374 \pm   0.09$  &     $ (7.193 \pm  0.596)\cdot10^{-3}$\\
  7473.60&	   I&       30x60&     $21.20   \pm  0.08$&    $(8.444  \pm  0.622)\cdot10^{-3}$ & $21.144 \pm   0.08$  &     $ (8.444 \pm  0.622)\cdot10^{-3}$\\
  9426.24&	   I&       30x60&     $20.66   \pm  0.05$&   $ (1.389  \pm  0.064)\cdot10^{-2}$ & $20.604  \pm  0.05$  &     $ (1.462 \pm 0.067)\cdot10^{-2}$\\
\hline
GROND&  &    &      &     & &  \\
\hline
20880.0	&  g$'$ &&	 $19.9  \pm    0.1$	&	$(3.98   \pm   0.37)\cdot10^{-2}$  &      $19.79\pm	0.1$	    &	$(4.406	\pm	0.410)\cdot10^{-2}$\\
20880.0	&  r$'$ &&	 $19.3  \pm    0.1$	&	$(6.92   \pm 0.64)\cdot10^{-2}$  &      $19.22\pm	0.1	$	&	$(7.454\pm		0.686)\cdot10^{-2}$\\
20880.0	&  i$'$ &&     $19.2  \pm    0.1$	&	$(7.59   \pm  0.70)\cdot10^{-2} $ &   	  $19.14\pm	0.1$    &	$(8.017	\pm	0.738)\cdot10^{-2}$\\
20880.0	&  z$'$ &&     $19.1  \pm    0.1$	&	$(8.38   \pm   0.77)\cdot10^{-2}$   &     $19.05\pm	0.1 $       &   $(8.694	\pm	0.801)\cdot10^{-2}$\\
20880.0	&  J  &&     $19.0  \pm    0.15$&	$(9.12   \pm  1.26)\cdot10^{-2}$  &      $18.97\pm	0.15$   &   $(9.359	\pm	1.293)\cdot10^{-2}$\\
20880.0	&  H  &&     $18.7  \pm    0.15$&	$(1.202   \pm  0.166 )\cdot10^{-1}$  &     $18.68	\pm0.15 $   &   $(1.221	\pm	0.169)\cdot10^{-1}$\\
20880.0	&  K  &&     $19.0  \pm    0.15$&	$(9.12   \pm  1.26 )\cdot10^{-2}$  &    $ 19.00\pm	0.15 $      &   $(9.135	\pm	0.502)\cdot10^{-2}$\\
112680. &  g$'$ &&     $21.26 \pm    0.05$&	$(1.14   \pm   0.05 )\cdot10^{-2}$  &     $21.15\pm	0.05  $     &   $(1.259	\pm	0.058)\cdot10^{-2}$\\
112680. &  r$'$ &&     $20.49  \pm   0.05$&	$(2.31   \pm  0.10)\cdot10^{-2}$	 & $20.41\pm	0.05$		& 	$(2.491\pm		0.115)\cdot10^{-2}$\\
112680. &  i$'$ &&     $20.24  \pm   0.05$&	$(2.91   \pm   0.13)\cdot10^{-2}$  &      $20.18\pm	0.05 $      &   $(3.076	\pm	0.142)\cdot10^{-2}$\\
112680. &  z$'$ &&     $19.99  \pm   0.05$&	$(3.66   \pm  0.17)\cdot10^{-2}$	&  $19.94\pm	0.05$       & 	$(3.830\pm		0.176)\cdot10^{-2}$\\
112680. &  J  &&     $19.6   \pm   0.1$	&	$(5.25   \pm  0.48)\cdot10^{-2}$	& $ 19.57	\pm0.1	$		&	$(5.386	\pm	0.496)\cdot10^{-2}$\\
\hline
PAIRITEL&  &    &      &     & &  \\
\hline
41133.2 &  J	&1875.67&	        $17.78\pm0.12$&	 $(1.232\pm 0.126)\cdot10^{-1}$	&   $17.752\pm0.12$ &       $(1.264 \pm  0.140)\cdot10^{-1}$\\
41133.2 &  H	&1875.67&	        $16.91\pm0.10$&  $(1.763\pm 0.162)\cdot10^{-1}$    & 	$16.893\pm0.10$ &       $(1.791  \pm 0.165)\cdot10^{-1}$\\
41133.2 &  $\rm{K_s}$	&1875.67&   $16.34\pm0.13$&  $(1.941\pm0.232)\cdot10^{-1}$     &   $16.3383\pm0.13$&       $(1.944 \pm  0.233)\cdot10^{-1}$\\
44006.0 & J & 1844.28&	           $17.60\pm0.11$ &  $(1.453  \pm 0.147)\cdot10^{-1}$&    $17.572\pm0.11$&       $(1.492 \pm 0.151)\cdot10^{-1}$\\
44006.0	& H	& 1844.28&	           $16.83\pm0.10$ &  $(1.898\pm0.174)\cdot10^{-1}$     &    $16.813\pm0.10$&       $(1.928\pm0.178)\cdot10^{-1}$\\
44006.0	& $\rm{K_s}$& 1844.28&    $15.87\pm0.10$  &  $(2.993  \pm0.276)\cdot10^{-1}$&  $15.8683\pm0.10$ &       $(2.993\pm0.276)\cdot10^{-1}$\\
\hline
NOT&  &    &      &     & &  \\
\hline
19680. & R	& & $19.23\pm0.03$ &  $ (6.255	 \pm0.200)\cdot10^{-2} $  &     $19.1545   \pm   0.03$&	  $(6.706 \pm 0.185)\cdot10^{-2}$\\
\hline
\end{tabular}
\caption{Ground-based photometric set of GRB\,081028.
Column 1: observations mid-time since BAT trigger;
column 2: photometric filter used; column 3: exposure;
columns 4 and 5: observed magnitude and flux; columns 6 and 7:
magnitudes and fluxes corrected for Galactic reddening. 
GROND data come from Clemens et al. (2008; 2008b).
CrAO data come from Rumyantsev et al., (2008). }
\label{Tab:OpticalData}
\end{center}
\end{table*}

\bsp

\label{lastpage}

\end{document}